\pdfoutput=1
\documentclass{article}

    \usepackage[final,nonatbib]{neurips_2024}

\usepackage[nonatbib]{neurips_2024}

\usepackage[table]{xcolor}
\usepackage[utf8]{inputenc} 
\usepackage[T1]{fontenc}    
\usepackage[colorlinks=true,linkcolor=blue,citecolor=blue]{hyperref}       
\usepackage{url}            
\usepackage{booktabs}       
\usepackage{amsfonts}       
\usepackage{nicefrac}       
\usepackage{microtype}      
\usepackage{xcolor}         
\usepackage[textsize=tiny]{todonotes}
\usepackage{amsmath}
\usepackage{amssymb}
\usepackage{mathtools}
\usepackage{amsthm}
\usepackage{graphicx}
\usepackage{subfigure}
\usepackage{multirow}
\usepackage{tablefootnote}
\usepackage{makecell}
\usepackage{algorithmic}
\usepackage[linesnumbered,ruled]{algorithm2e}
\usepackage{wrapfig}
\usepackage{lipsum,caption}
\usepackage{pifont}
\usepackage{lipsum}
\newtheorem{theorem}{Theorem}
\newtheorem{lemma}{Lemma}

\title{\method: Efficient Private Inference via \\ Block Circulant Transformation}

\newcommand{\method}{PrivCirNet}
\newcommand{\encode}{CirEncode}

\newcommand{\bm}{\boldsymbol}
\newcommand{\lj}{\left \langle}
\newcommand{\rj}{\right \rangle}

\newcommand{\Enc}{\mathrm{Enc}}

\definecolor{Gray}{gray}{0.85}
\author{%
  \makebox[0.3\linewidth]{Tianshi Xu}\\
  \makebox[0.3\linewidth]{Peking University}\\
  \makebox[0.3\linewidth]{\texttt{tianshixu@stu.pku.edu.cn}} \\
  \And
  \makebox[0.3\linewidth]{Lemeng Wu} \\
  \makebox[0.3\linewidth]{Meta, Inc.} \\
  \makebox[0.3\linewidth]{\texttt{lmwu@meta}} \\
  \AND
  \makebox[0.3\linewidth]{Runsheng Wang} \\
  \makebox[0.3\linewidth]{Peking University} \\
  \makebox[0.3\linewidth]{\texttt{r.wang@pku.edu.cn}} \\
  \And
  \makebox[0.3\linewidth]{Meng Li$^*$} \\
  \makebox[0.3\linewidth]{Peking University} \\
  \makebox[0.3\linewidth]{\texttt{meng.li@pku.edu.cn}} \\
}

\renewcommand{\thefootnote}{}
\begin{document}
\footnote{$^*$Corresponding author: meng.li@pku.edu.cn}
\renewcommand{\thefootnote}{\arabic{footnote}}
\setcounter{footnote}{0} 
\maketitle

\begin{abstract}
    Homomorphic encryption (HE)-based deep neural network (DNN) inference protects data and model privacy but suffers from significant computation overhead. We observe transforming the DNN weights into circulant matrices converts general matrix-vector multiplications into HE-friendly 1-dimensional convolutions, drastically reducing the HE computation cost. Hence, in this paper, we propose \method, a protocol/network co-optimization framework based on block circulant transformation. At the protocol level, \method~customizes the HE encoding algorithm that is fully compatible with the block circulant transformation and reduces the computation latency in proportion to the block size. At the network level, we propose a latency-aware formulation to search for the layer-wise block size assignment based on second-order information. \method~also leverages layer fusion to further reduce the inference cost. We compare \method~with the state-of-the-art HE-based framework Bolt (IEEE S\&P 2024) and HE-friendly pruning method SpENCNN (ICML 2023). For ResNet-18 and Vision Transformer (ViT) on Tiny ImageNet, \method~reduces latency by $5.0\times$ and $1.3\times$ with iso-accuracy over Bolt, respectively, and improves accuracy by $4.1\%$ and $12\%$ over SpENCNN, respectively. For MobileNetV2 on ImageNet, \method~achieves $1.7\times$ lower latency and $4.2\%$ better accuracy over Bolt and SpENCNN, respectively. 
    Our code and checkpoints are available on \href{https://github.com/Tianshi-Xu/PrivCirNet}{Git Hub}.

\end{abstract}

\vspace{-10pt}

\section{Introduction}
\label{sec:introduction}

The past few years have witnessed the rapid evolution of deep learning (DL) as well as its increasing adoption in sensitive and private applications, including face authentication~\cite{azouji2022efficientmask_face}, medical diagnosis~\cite{kaissis2021end_medical}, code auto-completion~\cite{xu2022systematic}, etc. Privacy emerges as a major concern and leads to a growing demand for privacy-preserving DL~\cite{Choi_Reagen_Wei_Brooks_Impala_2022,Gupta_Kumaraswamy_Chandran_Gupta_LLAMA_2022,rathee2020cryptflow2,pang2023bolt}. Homomorphic encryption (HE) is proposed as a promising technology for privacy protection and attracts a lot of attention~\cite{gilad2016cryptonets,Juvekar_Vaikuntanathan_gazelle_2018,huang2022cheetah,pang2023bolt}. By encrypting the data into ciphertexts, HE allows computation over the encrypted data directly and produces encrypted results, without leaking any knowledge of the data itself~\cite{gilad2016cryptonets}.

\begin{figure}[!tb]
    \centering
    \includegraphics[width=1.0\linewidth]{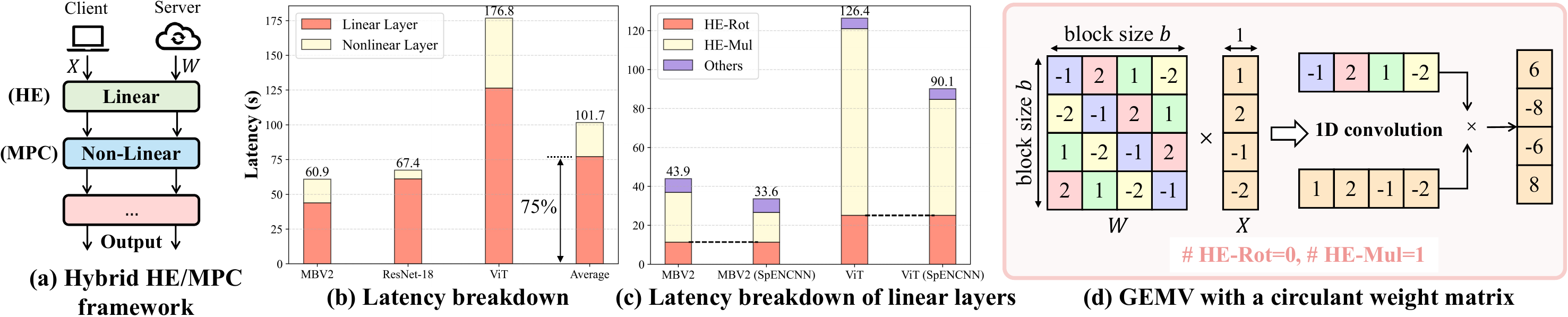}
    \caption{(a) Illustration of Hybrid HE/MPC-based private inference; (b) latency breakdown of linear layers and nonlinear layers based on Bolt's protocol;   (c) latency breakdown of linear layers of the original model and SpENCNN with 50\% sparsity; (d) GEMV with a circulant weight matrix.}
    \label{fig:intro}
\end{figure}

To apply HE for private deep neural network (DNN) inference, there are two main approaches, including the end-to-end HE-based schemes~\cite{gilad2016cryptonets,lou2019she,lee2022privacy,lee2022low,park2023toward_fhe,kim2023optimized_fhe,fan2023tensorfhe,kim2022ark,onoufriou2021fully} and the \textbf{hybrid HE/multi-party computation (MPC)-based schemes}~\cite{pang2023bolt,huang2022cheetah,Liu_Juuti_MiniONN_2017,Mishra_Delphi_2020,Garimella_Ghodsi_Jha_Garg_Reagen_2022,lu2023bumblebee,kim2023hyphen}. As shown in Figure~\ref{fig:intro} (a), the hybrid HE/MPC scheme leverages HE and MPC protocols to evaluate the linear and nonlinear layers separately, which usually demonstrates better accuracy due to its ability to realize accurate activation functions~\cite{hao2022iron}. In contrast, the end-to-end scheme relies on polynomial approximation or TFHE schemes for activation functions, which either suffer from low accuracy or low computation efficiency~\cite{lou2020falcon,lou2019she}. Hence,\textbf{ we focus on the hybrid scheme in our paper.}

While formal privacy protection can be achieved, HE-based DNN inference suffers from high computation cost and orders of magnitude latency overhead~\cite{pang2023bolt,huang2022cheetah}. Previous works have proposed algorithm-level optimizations on HE encoding and DNN architectures. HE encoding translates high-dimensional tensor operations of DNNs into 1-dimensional polynomial operations of HE and directly impacts the computation efficiency. For example, Cheetah~\cite{huang2022cheetah} and Falcon~\cite{xu2023falcon} propose efficient encoding algorithms for convolutions while Iron~\cite{hao2022iron} and BubbleBee~\cite{lu2023bumblebee} optimize for general matrix multiplications (GEMMs). Neujeans~\cite{ju2023neujeans} and Bolt~\cite{pang2023bolt} further introduce the baby-step giant-step (BSGS) algorithm to reduce the number of HE rotations and achieve state-of-the-art (SOTA) performance. While significant speedup has been achieved, the overall latency of MobileNetV2 \cite{sandler2018mobilenetv2} and Vision Transformer (ViT) \cite{dosovitskiy2020image} still exceeds 60s and 170s with Bolt, respectively, as shown in Figure~\ref{fig:intro} (b) and (c). Meanwhile, linear layers account for more than 75\% of total latency due to HE multiplications and rotations, thus, becoming the main optimization target of \method.



DNN model optimizations focus on developing HE-friendly architectures.~\cite{cho2022SNL,jha2021deepreduce,kundu2023SENet,cho2022sphynx,peng2023autorep,zeng2022mpcvit,kundu2023making} optimize the activation functions for communication and computation reduction, which is orthogonal to our work.~\cite{spencnn,Cai_Zhang_Ning_Xin_Wu_Hunter,HE-PEx} propose HE-friendly structured pruning to reduce both HE rotations and multiplications. However, as shown in Figure~\ref{fig:intro} (c), as these methods are not fully compatible with the SOTA protocols, their latency reduction remains limited, especially for HE rotations\footnote{The incompatibility is due to the BSGS algorithm and is explained in Appendix~\ref{app:spencnn_fail} in detail.}.

To further reduce the computation cost of linear layers and bridge the latency gap, in this paper, we propose~\method. \textit{Our key observation is that the circulant transformation of weight matrices enables to convert a general matrix-vector multiplication (GEMV) into a HE-friendly 1-dimensional convolution, simultaneously reducing the HE multiplications and rotations,} as shown in Figure~\ref{fig:intro} (d). While directly transforming the whole weight matrix into a circulant matrix incurs high accuracy degradation, we propose block circulant transformation and answer the following two questions. First, existing HE encoding algorithms are not fully compatible with block circulant weight matrices, limiting the efficiency gain. How to co-design the encoding algorithm to fully unleash the potential is the first question. Meanwhile, as block circulant transformation introduces structure constraints to weight matrices and inevitably impacts the accuracy, how to determine the layer-wise block sizes for better accuracy/efficiency trade-off becomes the second question. 

\method~features a novel encoding algorithm optimized for block circulant weight matrices, dubbed CirEncode, that reduces the HE computation in proportion to $block\ size$. \method~also co-design a latency-aware optimization formulation for layer-wise block size assignment based on second-order information.~\method~further leverages layer fusion to reduce the inference cost.
With extensive experiments across different DNN architectures (i.e., MobileNetV2, ResNet-18 and ViT) and datasets (i.e, CIFAR, Tiny ImageNet, and ImageNet), we demonstrate \method~reduces the latency of MobileNetV2, ResNet-18, and ViT by $1.7\times$, $5.0\times$ and $1.3\times$ compared with Bolt~\cite{pang2023bolt}, respectively. Compared with SOTA HE-friendly pruning method SpENCNN~\cite{spencnn}, \method~achieves $4.2$\%, $4.1$\%, and $12$\% better accuracy on MobileNetV2, ResNet-18, and ViT, respectively, demonstrating great capability to accelerate private inference for both ConvNets and Transformers.

\section{Preliminaries}
\label{sec:pre}

\textbf{Notations.} We represent matrices with upper-case letters (e.g., ${X}$) and vectors with lower-case letters (e.g., ${x}$). We also use lower-case letters with a ``hat'' symbol (e.g., $\hat{x}$) to represent a polynomial, and $\hat{x}[i]$ to denote the $i$-th coefficient of $\hat{x}$. We use $\times$ to represent polynomial multiplication and $\odot$ to denote element-wise multiplication. Let $\lceil \cdot \rceil$ denote ceiling operations and $[n]$ denote the set $\{0,\ldots, n-1\}$ for $n\in \mathbb{Z}^+$, where $\mathbb{Z}$ denotes the integer domain. We also denote the set of integer polynomials with $\mathbb{A}_{n}=\mathbb{Z}[X]/(X^n-1)$, whose degree $n$ is a power-of-two integer (e.g., $2^{13}$ following Bolt~\cite{pang2023bolt}). We use $(d_1,d_2,d_3)$ to denote the input, hidden, and output dimensions of a GEMM, respectively. For convolution, we use $(H, W, C)$ to represent the input height, width, and number of input channels, and $(R, K)$ to denote the kernel size and number of output channels.

\subsection{Cryptographic Primitives}


\textbf{BFV HE Scheme.} Following most hybrid HE/MPC schemes~\cite{pang2023bolt,gilad2016cryptonets,Juvekar_Vaikuntanathan_gazelle_2018,huang2022cheetah,Mishra_Delphi_2020}, \method~leverages the lattice-based Brakerski-Fan-Vercauteren (BFV) HE scheme~\cite{fan2012somewhat} and mainly involves the following HE operations, including ciphertext addition (denoted as HE-Add), ciphertext-plaintext multiplication (denoted as HE-Pmult), and ciphertext rotation (denoted as HE-Rot). While HE-Pmult and HE-Rot dominate the overall computation cost, each HE-Rot operation is usually an order of magnitude slower than HE-Pmult~\cite{spencnn,ran2024penguin}.

\textbf{HE Encoding Methods.} HE operates over polynomials with 1-dimensional coefficient vectors while DNNs compute over tensors. Encoding is the procedure to map a tensor to a polynomial and directly determines the computation efficiency. Existing encoding methods can be classified into two categories: coefficient encoding~\cite{huang2022cheetah,hao2022iron,xu2023falcon,lu2023bumblebee} and single instruction multiple data (SIMD) encoding~\cite{Juvekar_Vaikuntanathan_gazelle_2018,rathee2020cryptflow2,zhang2021gala,ju2023neujeans,pang2023bolt}. Coefficient encoding can support convolutions efficiently with a single HE-Pmult~\cite{huang2022cheetah}. In contrast, SIMD encoding only supports element-wise multiplications and requires multiple HE-Rot for convolutions~\cite{Juvekar_Vaikuntanathan_gazelle_2018}. For GEMMs, either coefficient encoding~\cite{lu2023bumblebee} or SIMD encoding~\cite{pang2023bolt} requires HE-Pmult and HE-Rot, 
while the SIMD encoding algorithm Bolt~\cite{pang2023bolt} achieves the SOTA computation efficiency.


The two encoding methods can be transformed to each other through the discrete Fourier transform (DFT) as shown in Lemma~\ref{lemma:DFT} \cite{ju2023neujeans}. The main reason is that polynomial multiplication implements convolutions in the coefficient domain and is equivalent to element-wise multiplications in the frequency domain, leading to Lemma~\ref{lemma:DFT} \cite{ju2023neujeans}. While \cite{ju2023neujeans} only leverages such nested encoding for convolutions, we show how such schemes can be improved to support block circulant GEMMs and convolutions. We refer interested readers to~\cite{ju2023neujeans} for a more detailed description.
\begin{lemma}
    \label{lemma:DFT}
    \resizebox{0.9\textwidth}{!}{
    $\lj \operatorname{DFT}(w) \rj_{\mathrm{SIMD}} \times \lj \operatorname{DFT}(x) \rj_{\mathrm{SIMD}} = \lj \operatorname{DFT}(w)\odot \operatorname{DFT}(x)\rj_{\mathrm{SIMD}}=\operatorname{DFT}(\lj w \rj_{\mathrm{Coeff}}\times \lj x \rj_{\mathrm{Coeff}})$
    }
\end{lemma}




\subsection{Threat Model and Security Guarantee}

\method~works in a general private inference scenario that involves two parties, i.e., server and client. A server holds the proprietary DNN model and a client owns private data \cite{huang2022cheetah,hao2022iron}. \method~enables the client to obtain the inference results while keeping the server's model weights and the client’s data private. Consistent with previous works~\cite{pang2023bolt,Juvekar_Vaikuntanathan_gazelle_2018,huang2022cheetah,hao2022iron}, we assume the DNN architecture (including the block sizes) is known to both sides and adopt an \textit{honest-but-curious} security model in which both parties follow the specification of the protocol but also try to learn more from than allowed. Following~\cite{pang2023bolt,huang2022cheetah},~\method~is built upon cryptographic primitives, including BFV and MPC protocols, and focuses on co-optimizing the DNN architecture and the HE encoding algorithm. The security can hence be guaranteed following~\cite{fan2012somewhat,goldreich1998secure}.



\subsection{Related Works}
\begin{table}[h]
\centering
\caption{Comparison with existing private inference works.
}
\label{tab:comp}
\resizebox{0.8\linewidth}{!}{
\Huge
\begin{tabular}{c|ccc|c|c}
\toprule
\multirow{2}{*}{Method} & \multicolumn{3}{c|}{HE Encoding Optimization}& \multirow{2}{*}{Target Ops} & \multirow{2}{*}{Network Optimization}    \\
\cmidrule{2-4}
& \makecell{Encoding} & \makecell{\# HE-Rot Reduction} &  \makecell{\# HE-Pmult Reduction} & & \\
\midrule
\cite{jha2021deepreduce,cho2022SNL,zeng2022mpcvit,cho2022sphynx} & \textcolor{red}{\ding{55}} & \textcolor{red}{\ding{55}} & \textcolor{red}{\ding{55}} &  ReLU/GELU  & ReLU/GELU Pruning    \\
\midrule
Cheetah~\cite{huang2022cheetah} &    Sparse      & \textcolor{green}{\ding{51}} & \textcolor{red}{\ding{55}}  & GEMV, Conv & /     \\
Iron~\cite{hao2022iron} &    Sparse      & \textcolor{green}{\ding{51}} & \textcolor{red}{\ding{55}}  & GEMM & /     \\
Neujeans~\cite{ju2023neujeans} &    Dense      & \textcolor{green}{\ding{51}} & \textcolor{red}{\ding{55}}  & Conv & /     \\
Bolt~\cite{pang2023bolt} &    Dense      & \textcolor{green}{\ding{51}} & \textcolor{red}{\ding{55}}  & GEMM & Token Pruning     \\
\cite{Cai_Zhang_Ning_Xin_Wu_Hunter,HE-PEx,spencnn} &    Dense      & \textcolor{red}{\ding{55}} & \textcolor{green}{\ding{51}}  & GEMM, Conv & Weight Pruning     \\
\midrule
\makecell{PrivCirNet (ours)}
 &    Dense      & \textcolor{green}{\ding{51}} & \textcolor{green}{\ding{51}}  & GEMM, Conv & \makecell{Block Circulant  Transformation}    \\
\bottomrule
\end{tabular}
}
\vspace{-5pt}
\end{table}
\begin{table}[h]
    \Huge
    \centering
    \caption{Comparison between~\method~and previous works that use circulant matrix.}
    \label{tab:cmp_cir}
    \resizebox{1.0\linewidth}{!}{
        \begin{tabular}{c|c|c|c|c|c|c}
        \toprule
        Method & Application & Initialization method & \makecell{Variable \\ block size} & Block size assignment & \makecell{Customized \\ Encoding Method} & Network \\
        \midrule 
        \makecell{CirCNN~\cite{ding2017circnn}\\ CirConv~\cite{liao2019circconv}}& Convolution in plaintext & Forbenius norm & \textcolor{green}{\ding{51}} & Uniform/Manually set& / & ConvNets   \\
        \midrule 
        Falcon~\cite{lou2020falcon} & End-to-end HE-based private inference & Forbenius norm & \textcolor{red}{\ding{55}} & Uniform& \textcolor{red}{\ding{55}} & Three-layer network   \\
        \midrule 
        \makecell{\method \\ (ours)} & Hybrid HE+MPC private inference & Loss-aware & \textcolor{green}{\ding{51}} & Latency-aware block size assignment& \textcolor{green}{\ding{51}} & ConvNets, Transformers   \\
        \bottomrule 
        \end{tabular}

    }
\end{table}
To improve the efficiency of HE-based DNN inference, existing works mainly focus on optimizing the HE encoding algorithm~\cite{huang2022cheetah,hao2022iron,xu2023falcon,Juvekar_Vaikuntanathan_gazelle_2018,rathee2020cryptflow2,zhang2021gala,ju2023neujeans,pang2023bolt} and the DNN architectures~\cite{jha2021deepreduce,cho2022SNL,kundu2023SENet,cho2022sphynx,peng2023autorep,zeng2022mpcvit,kundu2023making,Cai_Zhang_Ning_Xin_Wu_Hunter,HE-PEx,spencnn,lou2020falcon}. 
In Table~\ref{tab:comp}, we compare~\method~with prior-art works qualitatively. As can be observed,~\method~features network and encoding co-optimization to improve the efficiency of both GEMMs and convolutions. 

Attempts have been made to use the circulant matrix to accelerate inference in plaintext~\cite{ding2017circnn,liao2019circconv} and ciphertext~\cite{lou2020falcon} domains. However, two unresolved problems remain in both domains: 1) how to initialize circulant matrices, and 2) determining block sizes for each layer. As a result, it is hard for~\cite{ding2017circnn,liao2019circconv,lou2020falcon} to be applied to more efficient networks, e.g., MobileNetV2, Transformers, etc. Additionally, in the ciphertext domain, ~\cite{lou2020falcon}~cannot fully leverage block circulant matrices, resulting in limited or even increased latency. In contrast,~\method~maximizes the potential of block circulant matrices by customizing the HE encoding algorithm and proposing new initialization and block size assignment algorithms, achieving a superior accuracy-latency trade-off.
We give a comprehensive comparison between~\method~and~\cite{ding2017circnn,liao2019circconv,lou2020falcon} in Table~\ref{tab:cmp_cir}.
We leave a more detailed review of existing works in Appendix~\ref{app:related_work}.



\begin{figure}[h]
\begin{minipage}{0.47\textwidth}
    \centering
    \includegraphics[width=0.90\textwidth]{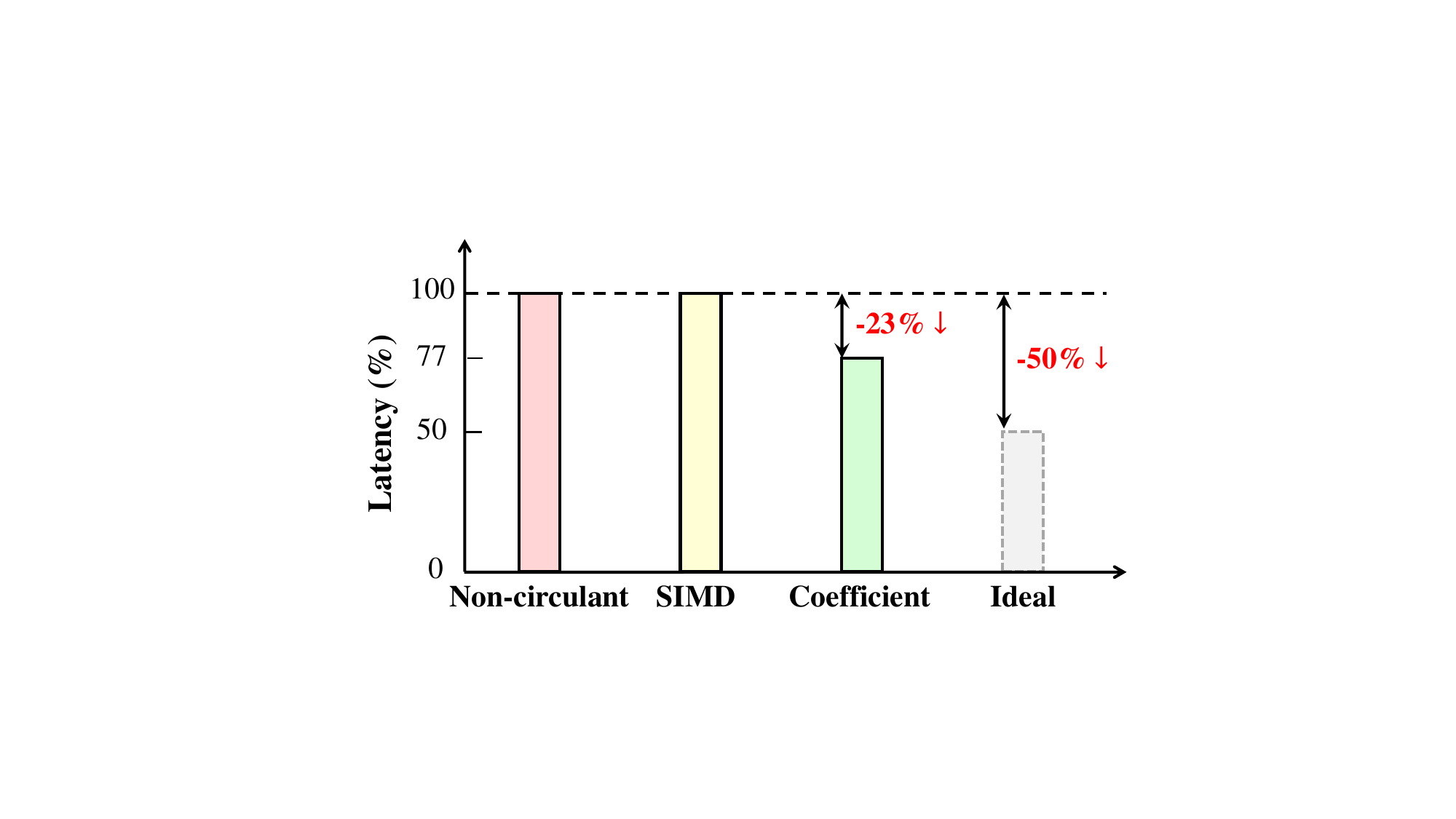} 
\end{minipage}
\hspace{3mm}
\begin{minipage}{0.49\textwidth}
    \Huge
    \centering
    \resizebox{0.8\linewidth}{!}{
    \begin{tabular}{c|c|c}
    \toprule 
    \toprule
    \makecell{Layer-wise block sizes} & Top-1 Acc. & Latency\\
    \midrule
    \rowcolor{gray!20}
    1-1-1-1 & 66.13 & 42 s \\
    16-16-16-1 & 64.51 & 25 s \\
    16-16-1-16 & 64.16 & 19 s \\
    16-1-16-16 & 63.23 & 16 s \\
    1-16-16-16 & 62.17 & 16 s \\
    \bottomrule
    \bottomrule
    \end{tabular}
    }
\end{minipage}\\
\begin{minipage}[t]{.47\textwidth}\centering
    \captionof{figure}{Directly using coefficient or SIMD encoding to block circulant GEMMs ($(d_1,d_2,d_3,b)=(256,192,576,2)$) leads to limited efficiency improvement.
    }
    \label{fig:observation1}
\end{minipage}\hspace{3mm}%
\begin{minipage}[t]{.49\textwidth}\centering%
    \captionof{table}{Accuracy and latency impact of applying block circulant transformation to different layers of MobileNetV2 on Tiny ImageNet. 32 layers are partitioned into 4 groups.}
    \label{tab:observation2}
\end{minipage}
\end{figure}
\section{\method~Framework}
\label{sec:method}

\subsection{Motivation}
\label{subsec:motiv}

While the circulant transformation enables to convert a GEMV into a HE-friendly 1-dimensional convolution, directly transforming the whole weight into a circulant matrix introduces large accuracy degradation due to the high compression ratio. 
We propose to leverage block circulant transformation and to trade off accuracy with efficiency by controlling the block sizes. However, we observe the following challenges that need to be addressed.

\textbf{Challenge 1: existing encoding algorithms are incompatible with block circulant weight matrices.} The computation of a GEMM with a block circulant weight matrix can be naturally decomposed into two steps, i.e., a circulant GEMV within each block and a general GEMM across blocks. Within each block, a circulant GEMV can be converted to a 1-dimensional convolution and be computed with a single HE-Pmult through coefficient encoding. However, when processing the GEMM across blocks, coefficient encoding suffers from either high communication cost~\cite{huang2022cheetah,hao2022iron} or extensive HE rotations~\cite{lu2023bumblebee}. In contrast, while SIMD encoding can process the GEMM across blocks more efficiently~\cite{pang2023bolt}, it still requires HE rotations to process the convolution within each block. As shown in Figure~\ref{fig:observation1}, with existing encoding algorithms, block circulant transformation only introduces limited efficiency improvement. \textit{Therefore, it is important to design the encoding algorithm to fully unleash the efficiency potential of the block circulant transformation.} 

\textbf{Challenge 2: accuracy and latency impact of block circulant transformation varies across layers.} We apply the block circulant weight transformation with different block sizes to different layers of a MobileNetV2 on Tiny ImageNet. As shown in Table~\ref{tab:observation2}, the accuracy and latency impact on the MobileNetV2 varies significantly. \textit{Hence, to better explore the Pareto optimal of efficiency and accuracy, layer-wise block size assignment becomes important}.

\begin{wrapfigure}[9]{r}[0em]{0.38\textwidth}
    \vspace{-5pt}
    \centering
    \includegraphics[width=0.90\linewidth]{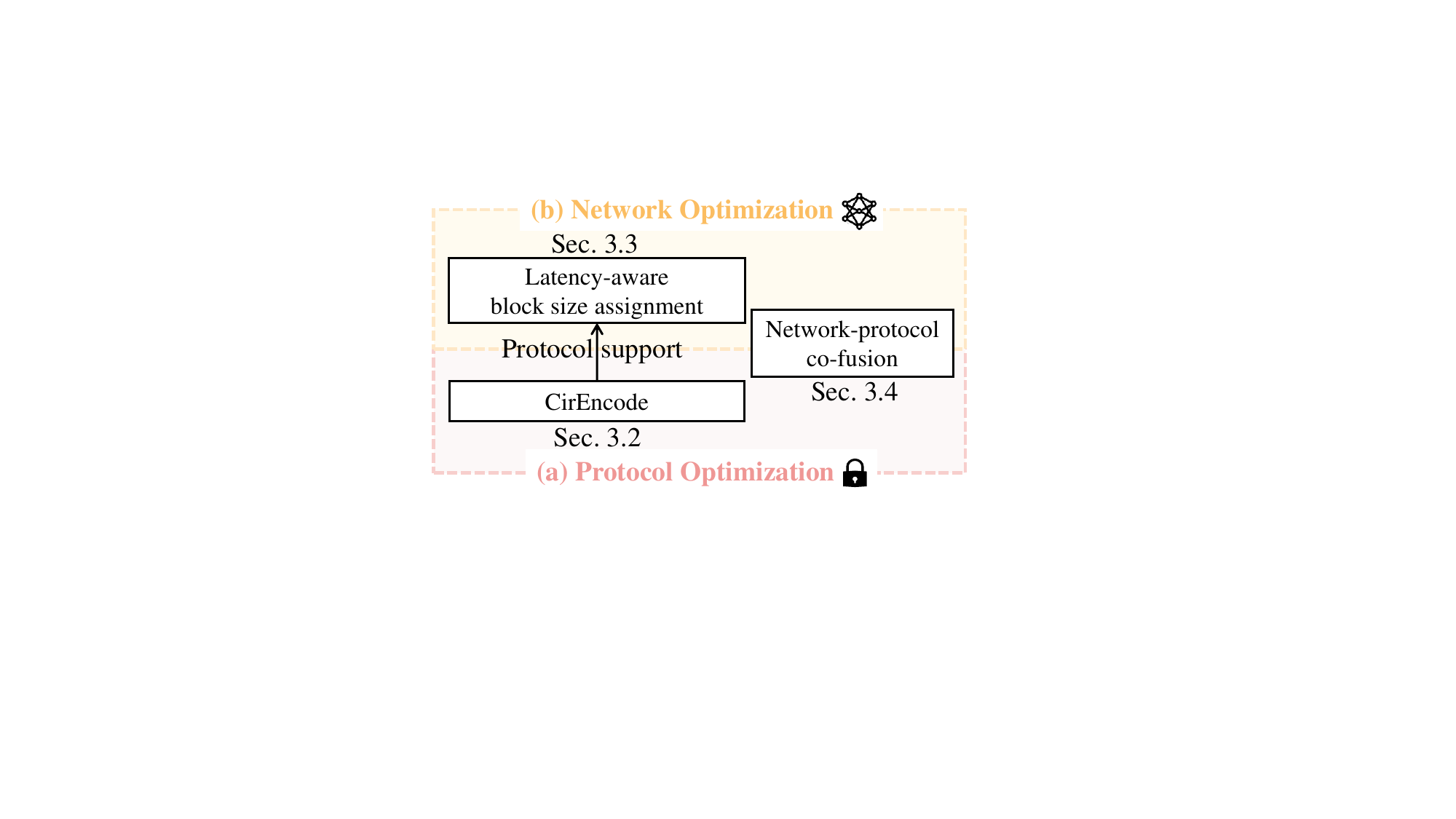}
    \caption{Overview of~\method.}
    \label{fig:overview}
\end{wrapfigure}
\textbf{\method~Overview.} In this paper, we introduce~\method, which features a joint optimization of the block circulant network and the private inference protocol. Figure~\ref{fig:overview} provides an overview of \method. We first propose \encode~for the GEMMs with block circulant weights in Section~\ref{sec:encode}. Then, we develop a latency-aware optimization algorithm to determine the block sizes for each layer based on second-order information in Section~\ref{sec:search}. We also propose network-protocol co-fusion methods to further boost the inference efficiency in Section~\ref{sec:fusion}.

\subsection{CirEncode: nested encoding for block circulant GEMMs}
\label{sec:encode}

\begin{figure}[!tb]
    \centering
    \includegraphics[width=1.0\linewidth]{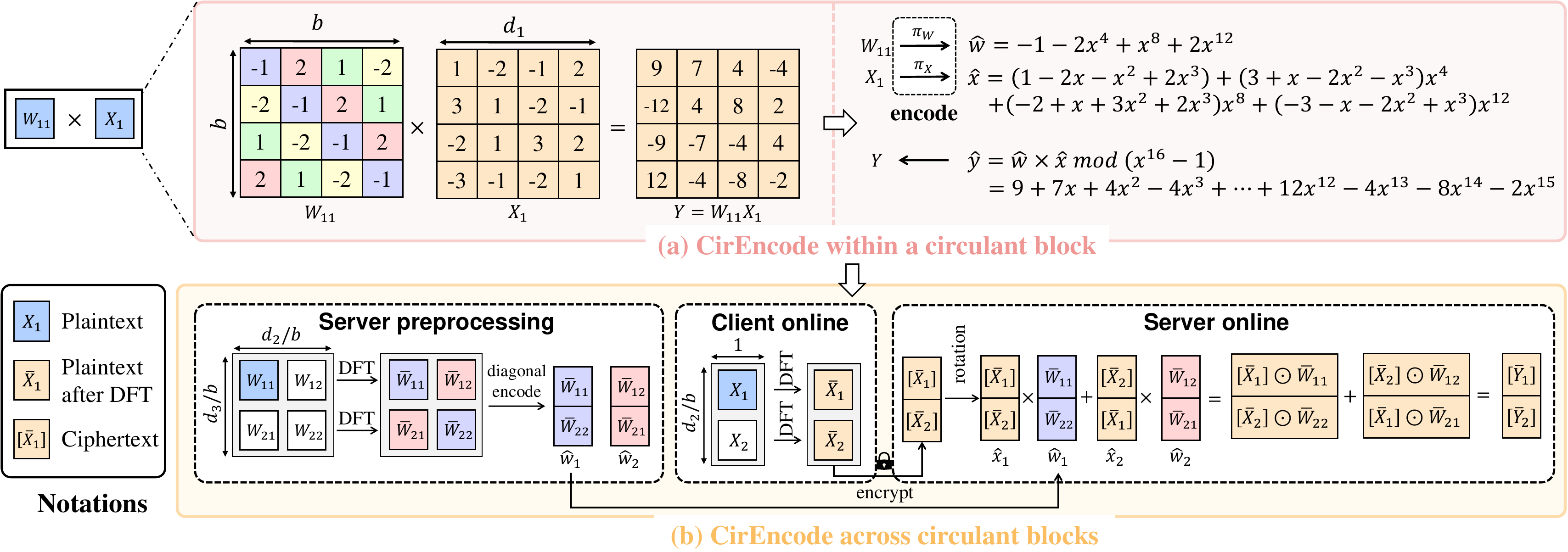}
    \caption{An example of~\encode~for block circulant GEMM where $(d_1,d_2,d_3,b)=(4,8,8,4)$.}
    \label{fig:cirencode}
\end{figure}


\textbf{High-level idea.} Consider a GEMM $Y = WX$, where $Y \in \mathbb{Z}^{d_3\times d_1}, W \in \mathbb{Z}^{d_3\times d_2}, X \in \mathbb{Z}^{d_2 \times d_1}$. $W$ is a block circulant matrix with block size $b$. Then, \encode~encodes the GEMM following two steps: for each block with $W \in \mathbb{Z}^{b \times b}$ and $X \in \mathbb{Z}^{b \times d_1}$, we convert the computation into $d_1$ parallel GEMVs and leverage the coefficient encoding to avoid HE-Rot as shown in Figure~\ref{fig:cirencode} (a); then, for across blocks, we regard it as a GEMM and leverage the SIMD encoding to further reduce the HE-Rot as shown in Figure~\ref{fig:cirencode} (b). Thereby, \encode~combines the advantages of both encoding schemes. 

\textbf{Encoding within a circulant block.} 
We elaborately design the encoding rule for a circulant GEMM. 
Formally, we define two encoding functions $\pi_\mathrm{W}: \mathbb{Z}^{b\times b}\rightarrow \mathbb{A}_n$ and $\pi_\mathrm{X}: \mathbb{Z}^{b\times d_1}\rightarrow \mathbb{A}_n$ as follows:
\begin{align*}
    \hat{w}=\pi_{\mathrm{W}}({W}), & \mathrm{\ \ where\ \ } \hat{w}[id_1]=W[i,0],\quad \forall i \in [b], j \in [d_1] \\
    \hat{x}=\pi_{\mathrm{X}}({X}), & \mathrm{\ \ where\ \ } \hat{x}[id_1+j]=X[i,j],\quad \forall i \in [b], j \in [d_1]
\end{align*}
where other coefficients of $\hat{w}$ are set to 0. $\hat{y}=\hat{w}\times \hat{x}$ directly gives the result of ${Y}={WX}$ as described in Theorem~\ref{theorem:encode} and we defer the proof to Appendix~\ref{proof:encode}.
\footnote{\encode~uses $\operatorname{mod} x^n-1$ which is different from~\cite{huang2022cheetah}, the explanation is in Appendix~\ref{proof:encode}.}
\begin{theorem}\label{theorem:encode}
    Given a circulant matrix $W \in \mathbb{Z}^{b \times b}$ and an input matrix $X \in \mathbb{Z}^{b \times d_1}$, where $bd_1 \leq n$, define two polynomials $\hat{w}=\pi_{\mathrm{W}}({W})$ and $\hat{x}=\pi_{\mathrm{X}}({X})$. Then, a GEMM ${Y}={WX}\in \mathbb{Z}^{b\times d_1}$ can be evaluated by the polynomial multiplication $\hat{y}=\hat{w}\times \hat{x}$, where ${Y}[i,j]=\hat{y}[id_1+j], \forall i \in [b], j \in [d_1]$.
\end{theorem}
Compared with prior-art coefficient encoding algorithms for a GEMM, e.g., Iron~\cite{hao2022iron},~\encode~features two key advantages: \textbf{\underline{1)}} the encoding density, i.e., number of useful elements encoded per polynomial, is much higher, minimizing the communication cost; \textbf{\underline{2)}} the input and output of a GEMM follow the same encoding rule described above, enabling layer fusion in Section~\ref{sec:fusion}. 



\textbf{Encoding across circulant blocks.}
Consider each circulant block as a unit, the computation across blocks can be regarded as a GEMM with dimension $(1,\frac{d_2}{b},\frac{d_3}{b})$. We apply the SIMD diagonal encoding to pack different circulant blocks in parallel and use DFT for each block to transform the coefficient encoding into the SIMD encoding format, as shown in Figure~\ref{fig:cirencode} (b). Similar to Lemma~\ref{lemma:DFT}, the correctness is given by Theorem~\ref{theorem:encode_across} and we defer the proof to Appendix~\ref{proof:encode_across}.
\begin{theorem}\label{theorem:encode_across}
    Given $M$ circulant weight matrices $W_0, \ldots, W_{M-1} \in \mathbb{Z}^{b\times b}$ and input matrices $X_0, \ldots, X_{M-1} \in \mathbb{Z}^{b\times d_1}$, define polynomials $\hat{w}_m$ and $\hat{x}_m$ with $m \in [M]$ following the coefficient packing in Theorem~\ref{theorem:encode}. Then, $Y_m = W_m X_m$ can be evaluated simultaneously through the polynomial multiplication in SIMD encoding: 
    \begin{align*}
    & \lj \operatorname{DFT}(\hat{y}_0)|\ldots|\operatorname{DFT}(\hat{y}_{M-1}) \rj_{\mathrm{Coeff}} \\ 
    = & \lj \operatorname{DFT}(\hat{w}_0)|\ldots|\operatorname{DFT}(\hat{w}_{M-1})\rj_{\mathrm{SIMD}} \times \lj \operatorname{DFT}(\hat{x}_0)|\ldots|\operatorname{DFT}(\hat{x}_{M-1})\rj_{\mathrm{SIMD}},
    \end{align*}
    where $|$ represents concatenation of polynomial coefficients and ${Y}_m[i, j]=\hat{y}_m[id_1+j], \forall i \in [b], j \in [d_1], m \in [M]$.
\end{theorem}
We further extend the BSGS algorithm~\cite{pang2023bolt} to~\encode~with details in Appendix~\ref{app:cirencode}. We also design~\encode~for block circulant convolutions as described in Appendix~\ref{app:encode_conv}.


\textbf{Theoretical complexity analysis.}
\begin{table}[!tb]
    \renewcommand{\arraystretch}{1.0}\
    \Huge
    \centering
    \caption{Theoretical complexity comparison of~\encode~with prior works. The data of GEMM is measured with dimension $(d_1,d_2,d_3)=(512,768,3072)$, and that of convolution is $(H,W,C,K,R)=(16,16,128,128,3)$. The polynomial degree $n=8192$ and block size $b=8$.}
    \label{tab:complexity_compare}
    \resizebox{\linewidth}{!}{
    \begin{tabular}{c|ccc|ccc}
    \toprule 
    \multirow{2}{*}{Framework} &  \multicolumn{3}{c|}{GEMM} & \multicolumn{3}{c}{Convolution} \\
    \cmidrule{2-7}
    &  \multicolumn{1}{c}{\# HE-Pmult}&\# HE-Rot & \# Ciphertexts & \multicolumn{1}{c}{\# HE-Pmult}&\# HE-Rot & \# Ciphertexts\\
    \midrule
    \multirow{2}{*}{CrypTFlow2~\cite{rathee2020cryptflow2}}& $O(d_1d_2d_3/n)$ &$O(d_1(d_2+d_3)/n+d_3)$ & $O(d_1(d_2+d_3)/n)$ & $O(HWCK/n)$ & $O(HW(C+K)/n+K)$ & $O(HW(C+K)/n)$\\
     &147456&3312&240 &9216 &208 &16 \\
    \midrule
    \multirow{2}{*}{Cheetah~\cite{huang2022cheetah}}& $O(d_1d_2d_3/n)$ &\multirow{2}{*}{0} &$O(d_1d_2/n+\lceil d_1/n\rceil d_3)$ &$O(HWCK/n)$ & \multirow{2}{*}{0} & $O(HWC/n+\lceil HW/n\rceil K)$  \\
    & 147456 & &3120 &1408 & &134\\
    \midrule
    \multirow{2}{*}{Iron~\cite{hao2022iron}}& $O(d_1d_2d_3/n)$ &\multirow{2}{*}{0} &$O(\sqrt{d_1d_2d_3/n})$ & $O(HWCKR^2/n)$ &\multirow{2}{*}{0} & $O(\sqrt{HWCKR^2/n})$ \\
    & 147456 & &960 &12672 & &257\\
    \midrule
    \multirow{2}{*}{Bumblebee~\cite{lu2023bumblebee}}& $O(d_1d_2d_3/n)$ &$O(d_1d_3\log_2n/(2\sqrt{n}))$ & $O(d_1(d_2+d_3)/n)$ &$O(HWCK/n)$ & $O(HWK\log_2n/(2\sqrt{n}))$ & $O(HW(C+K)/n)$\\
    & 147456 & 6144 &240& 1408& 256& 16 \\
    \midrule
    \multirow{2}{*}{Neujeans+BSGS~\cite{ju2023neujeans}}& $O(d_1d_2d_3/n)$ &$O(\sqrt{d_1d_2d_3/n})$  &$O(d_1(d_2+d_3)/n)$ &$O(HWCK/n)$ & $O(\sqrt{HWCK/n})$ & $O(HW(C+K)/n)$ \\
    & 147456 & 588 &240 &1024 &48 &16  \\
    \midrule
    \multirow{2}{*}{Bolt+BSGS~\cite{pang2023bolt}}& $O(d_1d_2d_3/n)$ &$O(\sqrt{d_1d_2d_3/n})$  &$O(d_1(d_2+d_3)/n)$ &$O(HWCKR^2/n)$ & $O(\sqrt{HWCKR^2/n})$ & $O(HW(CR^2+K)/n)$ \\
    & 147456 & 528 &240 &11700 &106 &100\\
    \midrule
    \multirow{3}{*}{\encode~(ours)}&\multicolumn{3}{c|}{GEMM with circulant weight matrix} & \multicolumn{3}{c}{Convolution with circulant weight kernel}\\
    \cmidrule{2-7}
    & $O(d_1d_2d_3/(nb))$ &$O(\sqrt{d_1d_2d_3/(nb)})$  &$O(d_1(d_2+d_3)/n)$&$O(HWCK/(nb))$ & $O(\sqrt{HWCK/(nb)})$ & $O(HW(C+K)/n)$\\
    &18432& 48 & 240 &128 &8 &16 \\
    \bottomrule
    \end{tabular}
    }
\end{table}
Table~\ref{tab:complexity_compare} shows the theoretical complexity comparison of~\encode~with prior-art encoding methods in the number of HE-Pmult, HE-Rot, and ciphertexts.~\encode~computes a $(d_1,b,b)$ circulant GEMM with only $O(bd_1/n)$ HE-Pmult and 0 HE-Rot. Therefore, compared to the SOTA scheme, i.e., Bolt and Neujeans,~\encode~reduces the number of HE-Pmult and HE-Rot by a factor of $b$ and $\sqrt{b}$, respectively, for both GEMM and convolution. A detailed proof of theoretical complexity is available in Appendix~\ref{app:cirencode}.

\subsection{Latency-aware block size assignment with loss-aware initialization}
\label{sec:search}
Previous works use uniform block size~\cite{ding2017circnn,lou2020falcon} or manually set the block sizes~\cite{liao2019circconv} for each layer, resulting in sub-optimal performance. We now propose a novel latency-aware block size assignment algorithm based on second-order information together with loss-aware initialization, which achieves a superior Pareto front of latency and accuracy.
 

\textbf{Loss-aware initialization for circulant matrices.}
Previously, circulant matrices were initialized by minimizing the Frobenius norm between the non-circulant and circulant matrices~\cite{Chu_Plemmons_2003,liao2019circconv}, i.e., $\min \left \|W_i^\prime-W_i\right\|^2_2$, where $W_i^\prime$ represents the weight matrix after the circulant transformation of layer $i$. While this method minimizes the min square error (MSE) of the weight matrix, it overlooks that \textit{the network accuracy has different sensitivity towards the MSE of different layers.} Therefore, we propose to directly assess the final loss instead of MSE for the transformation with Taylor expansion:
\begin{equation}
    \mathcal{L}_{W_i^\prime}(\mathcal{D})-\mathcal{L}_{W_i}(\mathcal{D})={\frac{\partial \mathcal{L}^{\top}(\mathcal{D})}{\partial W_i} \Delta W_i}+\frac{1}{2} \Delta W_i^{\top} H \Delta W_i+\mathcal{O}\left(\left\|\Delta W_i\right\|^3\right),
\end{equation}
where $\mathcal{L}$ is the task loss, $\mathcal{D}$ is the training dataset, $H$ is the Hessian matrix and $\Delta W_i=W_i^\prime-W_i$. The first term can be neglected as the model converges on the training dataset~\cite{li2021brecq}. The Hessian matrix can be approximated using diagonal Fisher information matrix~\cite{Montavon_Orr_Mller_2012}. We then define the sensitivity of layer $i$ as $\Omega_i$:
\begin{equation}
    \Omega_i = \Delta W_i^{\top} H \Delta W_i \approx \Delta W_i^{\top} \operatorname{diag}\left(\left(\frac{\partial \mathcal{L}(\mathcal{D})}{\partial W_i}\right)^2 \right)  \Delta W_i 
\end{equation} 
Hence, we propose initializing the circulant matrix by minimizing $\Omega_i$ instead of the Frobenius norm, which can be solved analytically as $W_i^\prime = \mathbb{E}\left[W_i \odot \left(\frac{\partial \mathcal{L}(\mathcal{D})}{\partial W_i}\right)^2 \right]_{diag}$. $\mathbb{E}_{diag}$ is the expectation of each diagonal of a matrix. An example is provided in Appendix~\ref{app:initialization}.

\textbf{Latency-aware block size assignment.}
Given an $L$-layer network, we denote the block size of each layer as $\{b_1,\ldots,b_L\}$, where $b_i\in \{2^0,\ldots,2^{k-1}\}$. The search space contains $k^L$ possible architectures, which can be extremely large, e.g., $2\times 10^{22}$ for $k=5,L=32$, rendering exhaustive search impractical. Therefore, we propose to formulate the search problem as an integer linear programming (ILP) problem, aiming to minimize the overall network sensitivity under the latency constraint~\cite{dong2019hawq,dong2020hawq,yao2021hawq}:
\begin{equation}
    \text{Objective: } \min_{\{ b_i \}^L_{i=1}} \quad \sum_{i=1}^L \Omega_{i}^{b_i} ,\quad
       \text{Subject to: } \quad \sum_{i=1}^L \operatorname{LAT}_{i}^{b_i} \le \text{Latency Limit}  
\end{equation}
Here, $\Omega_{i}^{b_i}$ is the $i$-th layer's sensitivity with block size $b_i$, $\operatorname{LAT}_{i}^{b_i}$ is the associated latency in private inference. $\operatorname{LAT}_{i}^{b_i}$ can be pre-computed given the dimension of the layer.
\begin{figure}[!tb]
    \centering
    \includegraphics[width=1.0\linewidth]{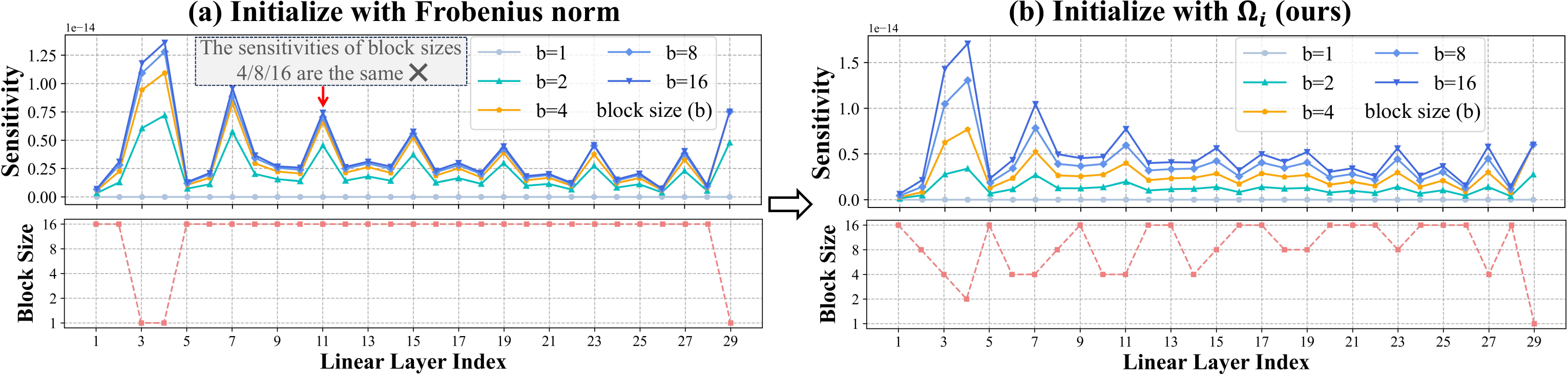}
    \caption{Layer-wise sensitivity and block size visualization for ViT on CIFAR-100.}\label{fig:sensitivity}
\end{figure}

\textbf{Visualization analysis.} 
We visualize the layer-wise sensitivity and the searched structure of different initialization methods in Figure~\ref{fig:sensitivity}. As we can observe in Figure~\ref{fig:sensitivity} (a), the previous method fails to tell the different sensitivity of block sizes 4, 8, and 16 for most of the layers. 
In contrast, our method, depicted in Figure~\ref{fig:sensitivity} (b), better captures the effects of varying block sizes on task loss.

\subsection{Network-Protocol Co-Fusion}\label{sec:fusion}
\begin{wrapfigure}[13]{r}[0em]{0.42\textwidth}
    \vspace{-10pt}
    \centering
    \includegraphics[width=0.90\linewidth]{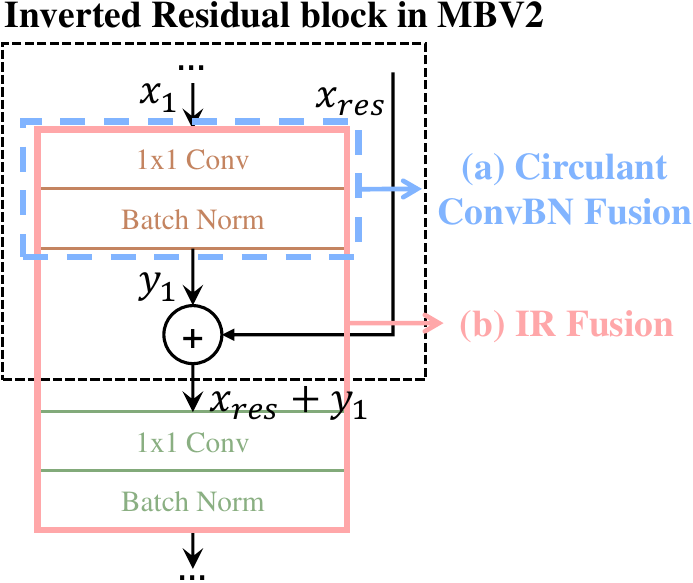}
    \caption{Network-Protocol Co-Fusion.}\label{fig:fusion}
\end{wrapfigure}
\textbf{Circulant ConvBN Fusion.} 
During the inference, convolution ($\operatorname{conv}$) and batch normalization ($\operatorname{bn}$) layers are usually fused for lower latency. However, na\"ive fusion destroys the block circulant structure. 
Hence, we propose a fusion method for circulant $\operatorname{conv}$ and $\operatorname{bn}$. Consider the learnable scaling factor $\gamma \in \mathbb{Z}^{C}$ for a $\operatorname{bn}$ layer. We combine the elements of $\gamma$ into groups of size $b$ and set $\gamma^\prime \in \mathbb{Z}^{C}$ such that $\gamma^\prime[i]=\frac{\sum_{j=0}^{b-1}\gamma [i+j-(i\mod b)]}{b}, \forall i\in [C]$.
We use the same strategy for the learnable bias, running mean, and variance, which maintains the circulant structure after fusion.

\textbf{Inverted Residual (IR) Fusion Protocol.} 
In the hybrid HE/MPC-based DNN inference, the network is evaluated layer by layer. We identify the potential for layer fusion of consecutive linear layers in MobileNetV2~\cite{sandler2018mobilenetv2}. Figure~\ref{fig:fusion} (b) shows where we implement fusion, aiming to compute $\operatorname{convbn}(x_{res} + \operatorname{convbn}(x_1))$ all together. 
Thanks to the encoding consistency provided by~\encode, we can fuse layers with equal block size. Details of the fusion algorithm are in Appendix~\ref{app:inverted_residual_fusion}.

\section{Experiments
}\label{sec:experiment}

\subsection{Experimental Setup}

\textbf{Implementation.}~\method~is built on top of the SEAL library \cite{sealcrypto} in C++. We use the OpenCheetah~\cite{huang2022cheetah} to evaluate Cheetah~\cite{huang2022cheetah} and CrypTFlow2~\cite{rathee2020cryptflow2}. We also implement Falcon~\cite{xu2023falcon}, Neujeans~\cite{ju2023neujeans} and Bolt~\cite{pang2023bolt} protocols.
Following~\cite{huang2022cheetah,Shen_Dong_Fang_Shi_Wang_Pan_Shi_ABNN2_2022,mohassel2017secureml}, we simulate a LAN network setting via Linux Traffic Control, where the bandwidth is 384 MBps and the echo latency is 0.3ms. All the experiments are performed on a machine with a 2.4 GHz Intel Xeon CPU. Following~\cite{pang2023bolt}, we set $n=8192$, security parameter $\lambda=128$, plaintext bitwidth $p=41$ and ciphertext bitwidth $q=218$, which is also the default setting in SEAL library~\cite{sealcrypto}.

\textbf{Datasets and Models.} We evaluate~\method~on MobileNetV2~\cite{sandler2018mobilenetv2}, ResNet-18~\cite{he2016resnet}, and ViT~\cite{dosovitskiy2020image} across four datasets: CIFAR-10, CIFAR-100, Tiny ImageNet and ImageNet.\footnote{Each of the models in the paper is capable of only classifying to the ImageNet 1k categories.} Detailed model architectures and training settings can be found in Appendix~\ref{app:exp_details}.

\textbf{Baselines.} We compare~\method~with prior-art HE-based DNN inference frameworks, including CrypTFlow2~\cite{rathee2020cryptflow2}, Cheetah~\cite{huang2022cheetah}, Falcon~\cite{xu2023falcon}, Neujeans~\cite{ju2023neujeans} and Bolt~\cite{pang2023bolt}. We also compare with SpENCNN~\cite{spencnn} which is the SOTA HE-friendly pruning method.


\subsection{Micro-Benchmark on Single GEMM and Convolution}\label{sec:exp_micro}
\textbf{Latency comparison.}
\begin{figure}[!tb]
    \centering
    \includegraphics[width=1.0\linewidth]{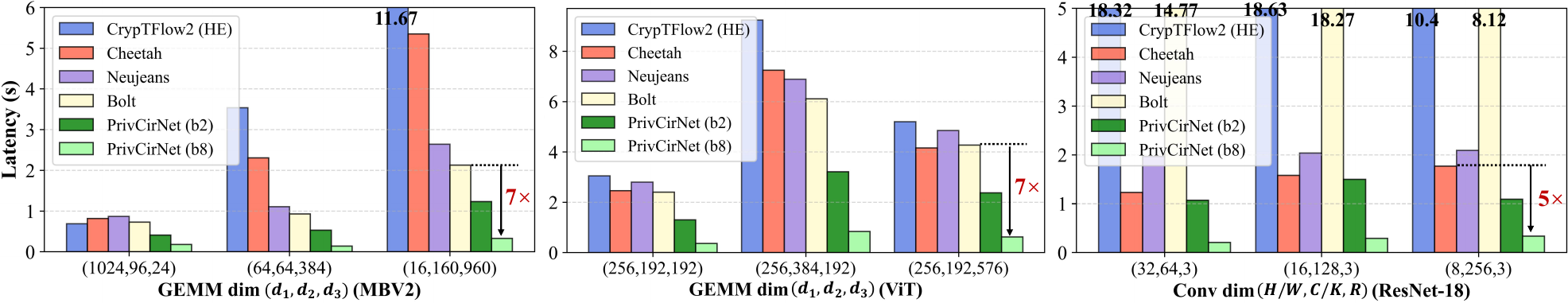}
    \caption{Latency comparison of different protocols for GEMMs and convolutions.~\method~use circulant weight with block size $b$.}
    \label{fig:exp_micro}
\end{figure}
In Figure~\ref{fig:exp_micro}, we benchmark~\method~on both GEMMs and convolutions with different block sizes. The layer dimensions are chosen from MobileNetV2, ResNet-18, and ViT. It can be observed that~\method~supports both GEMMs and convolutions efficiently. Compared with Bolt and Cheetah,~\method~(b8), i.e., block size of 8, achieves $5\sim 7\times$ latency reduction. With~\method~(b2), we can reduce latency by $1.7\times$ on average.

\textbf{The number of HE-Pmult and HE-Rot comparison. }\label{app:exp_rot_mul_cmp}
In Table~\ref{tab:exp_micro_rot}, we show the number of HE-Rot and HE-Pmult comparisons with different protocols. The layer dimensions are chosen from MobileNetV2, ResNet-18, and ViT. It can be observed that: \textbf{\underline{1)}} Compared with SOTA algorithms Bolt and Neujeans,~\method~(b8) achieves on average $7\times $ HE-Rot reduction and $8.5\times$ HE-Pmult reduction. And~\method~(b2) achieves on average $2.1\times $ HE-Rot reduction and $1.9\times$ HE-Pmult reduction which is consistent with the theoretical complexity. \textbf{\underline{2)}}~\method~supports both GEMM and convolution efficiently. On the contrary, Neujeans performs worse in GEMM while Bolt performs worse in convolution. 

\begin{table}[!tb]
    \renewcommand{\arraystretch}{1.0}
    \centering
    \Huge
    \caption{The number of HE-Rot / HE-Pmult comparisons of different protocols for GEMMs and convolutions with different dimensions.}
    \label{tab:exp_micro_rot}
    \resizebox{1.0\linewidth}{!}{
    \begin{tabular}{c|ccc|ccc|ccc|c}
    \toprule 
    \multirow{2}{*}{Method}&\multicolumn{3}{c|}{MobileNetV2}&\multicolumn{3}{c|}{ViT} &\multicolumn{3}{c|}{ResNet-18} &\multirow{2}{*}{Average}  \\
    \cmidrule{2-10}
    & (1024,96,24) &  (64,64,384)&(16,160,960)&(256,192,192)&(256,192,576)&(256,384,192)&(32,64,3)&(16,128,3)&(8,256,3)\\
    \midrule
    Neujeans+BSGS~\cite{ju2023neujeans} & 32 / 288&44 / 384&88 / 1024&90 / 1152&150 / 3456&120 / 2304&32 / 1024&48 / 1024&42 / 1134& 72 / 1310 \\
    Bolt+BSGS~\cite{pang2023bolt} &21 / 288 &33 / 384&55 / 960&60 / 1152&94 / 3456&78 / 2304&63 / 9216&106 / 11700&116 / 4608& 70 / 2504 \\
    \method~(b2)&9 / 144&21 / 192&37 / 480&36 / 576&60 / 1728&54 / 1152&16 / 512&32 / 726&28 / 567& \textbf{33 / 675} \\
    \method~(b8)&0 / 36&7 / 48&15 / 120&12 / 144&18 / 432&18 / 288&0 / 64&8 / 128&12 / 135& \textbf{10 / 155 } \\
    \bottomrule
    \end{tabular}
    }
\end{table}

\subsection{End-to-End Inference Evaluation}

\begin{figure}[!tb]
    \centering
    \includegraphics[width=\linewidth]{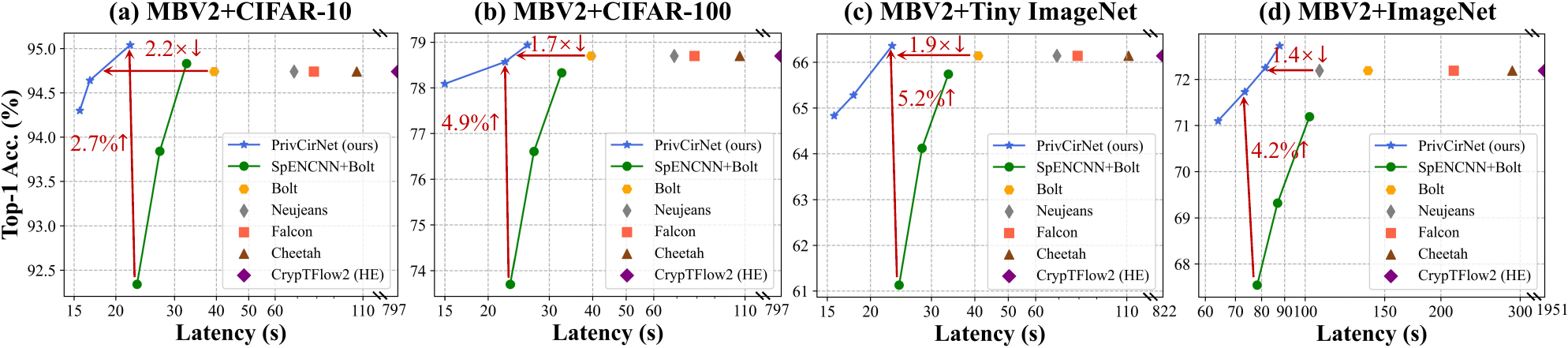}
    \caption{Comparison with SpENCNN and other prior-art protocols on MobileNetV2. 
    }
    \label{fig:exp_mbv2}
\end{figure}

\begin{figure}[!tb]
    \centering
    \includegraphics[width=\linewidth]{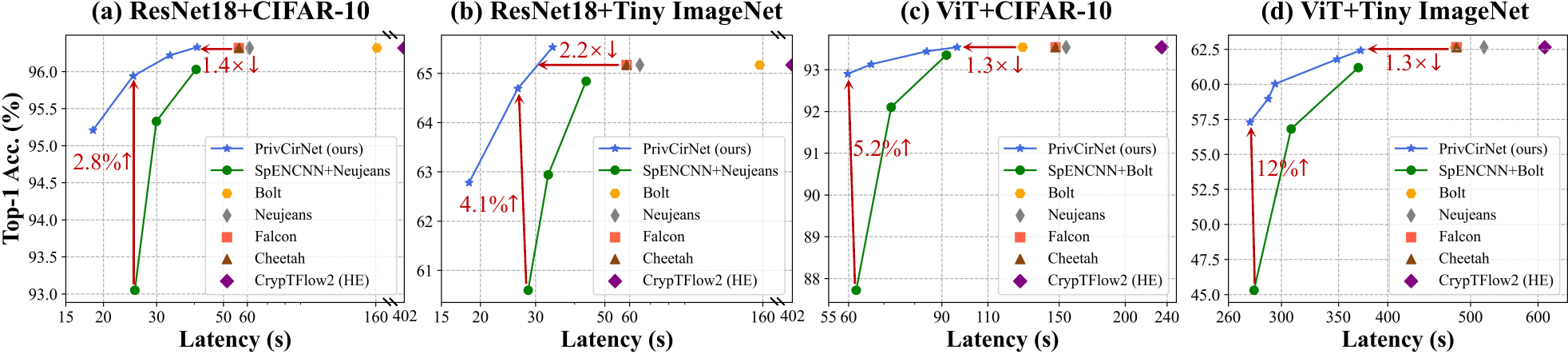}
    \caption{Comparison with SpENCNN and other prior-art protocols on ResNet-18 and ViT.} 
    \label{fig:exp_vit_res}
\end{figure}

In Figure~\ref{fig:exp_mbv2} and Figure~\ref{fig:exp_vit_res}, we benchmark~\method~at the full network scale and plot the Pareto front of accuracy and \textbf{latency of linear layers}. We make the following observation:

\textbf{Comparison with prior-art HE-based frameworks.}~\method~consistently outperforms prior-art frameworks, including Bolt, Neujeans, Falcon, etc, in both ConvNets and Transformers. Specifically, on Tiny ImageNet, compared with Bolt,~\method~achieves $1.9\times, 5.0\times, 1.3\times$ latency reduction with iso-accuracy on MobileNetV2, ResNet-18, and ViT, respectively. Compared to Cheetah,~\method~achieves $1.3\sim 4.8\times$ latency reduction with iso-accuracy across three models.

\textbf{Comparison with prior-art structured pruning method SpENCNN.}~\method~achieves SOTA accuracy/latency Pareto front across different datasets and models. Especially in larger compression ratios, SpENCNN suffers from huge accuracy loss. In comparison,~\method~outperforms SpENCNN by $5.2\%$ on MobileNetV2, $4.1\%$ on ResNet-18, and $12\%$ on ViT on Tiny ImageNet.

\textbf{Benchmark on ImageNet.} We benchmark~\method~on ImageNet with MobileNetV2 in Figure~\ref{fig:exp_mbv2} (d).~\method~achieves $1.4\times$ latency reduction compared with prior-art framework Neujeans and achieves $4.2\%$ accuracy improvement over SpENCNN with lower latency.


\subsection{Ablation Study}
\begin{wrapfigure}[12]{r}[0em]{0.46\textwidth}
    \vspace{-5pt}
    \centering
    \includegraphics[width=1.0\linewidth]{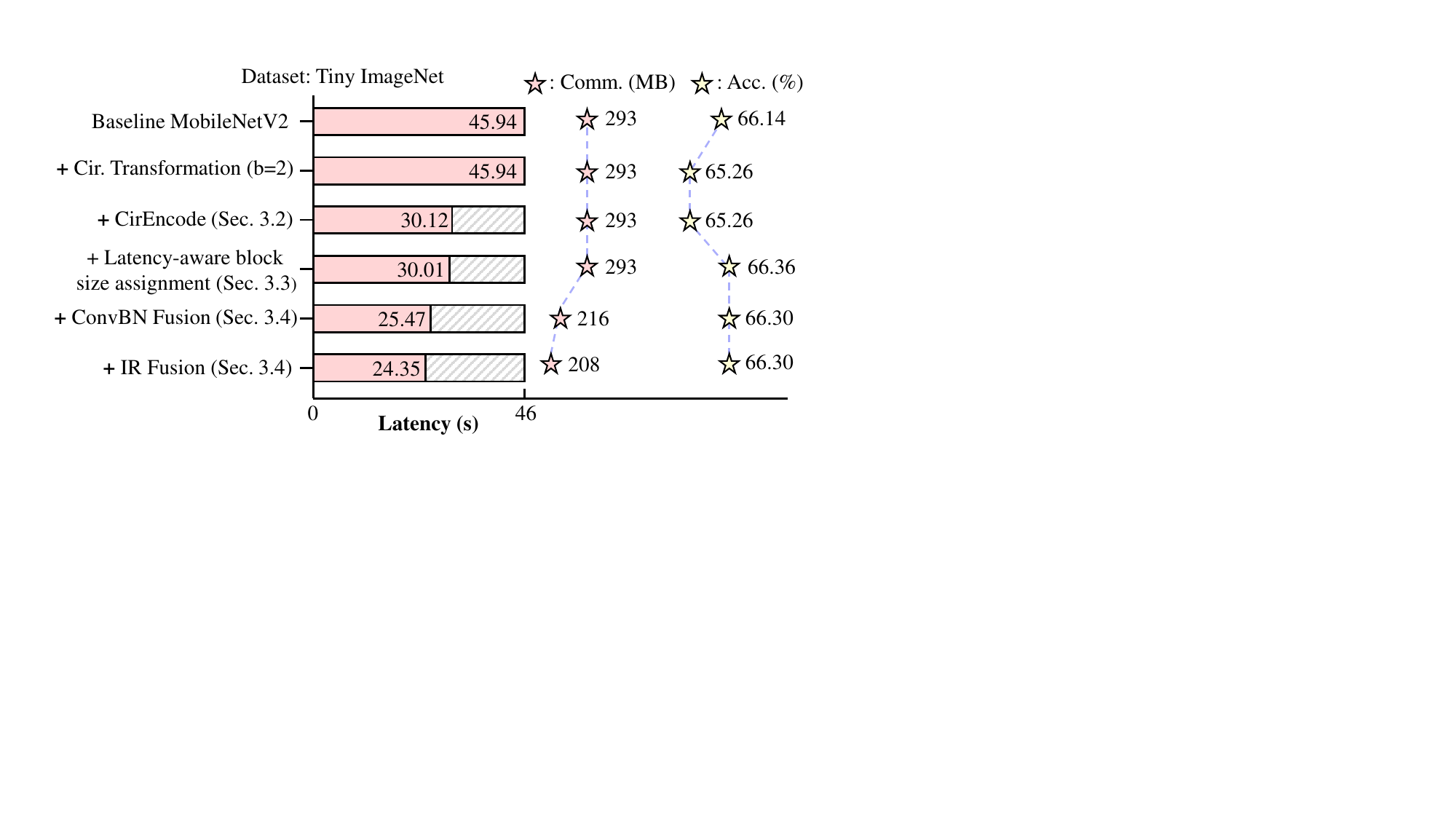}
    \caption{Ablation study of our proposed optimizations in~\method~on MobileNetV2.}\label{fig:exp_ablation}
\end{wrapfigure}
\textbf{Effectiveness of latency-aware block size assignment.} Table~\ref{tab:ablation_acc} shows the comparison of different block size assignment methods, including uniform block size, mixed block sizes with Frobenius norm initialization \cite{Chu_Plemmons_2003,liao2019circconv}, and mixed block sizes with loss-aware initialization. According to the results, we find that: \textbf{\underline{1)}}~\method~achieves the highest accuracy across most datasets and models. \textbf{\underline{2)}}~\method~exhibits enhanced performance at higher compression ratios, emphasizing the importance of latency-aware block size assignment in networks with limited capacity.

\textbf{Effectiveness of different optimizations in~\method.} We demonstrate the effectiveness of the proposed optimizations by adding them step by step on MobileNetV2 and Tiny ImageNet. As in Figure~\ref{fig:exp_ablation}, we observe that: \textbf{\underline{1)}} without~\encode, circulant transformation harms the accuracy and cannot reduce latency due to incompatibility with existing encoding algorithms; \textbf{\underline{2)}} latency-aware block size assignment significantly improves the accuracy and even outperforms the uncompressed model; \textbf{\underline{3)}} the fusion methods reduce both the latency and communication with negligible accuracy loss.

\textbf{Additional Results.} We present extra experiments to show 1) latency breakdown, and 2) comparison on more networks in Appendix~\ref{app:add_exp}.

\begin{table}[!tb]
    \renewcommand{\arraystretch}{1.1}
    \Huge
    \centering
    \caption{Accuracy comparison of different block size assignment methods. Latency limitation represents the proportion of latency relative to the original uncompressed model.}
    \label{tab:ablation_acc}
    \resizebox{0.9\linewidth}{!}{
    \begin{tabular}{c|c|c|c|c|c|c|c}
    \toprule 
    \multirow{3}{*}{Method}& \multirow{3}{*}{\makecell{Latency \\ Limitation}} & \multicolumn{6}{c}{Top-1 Acc. $\uparrow$} \\
    \cline{3-8}
    && \multicolumn{3}{c|}{MobileNetV2} & \multicolumn{3}{c}{ViT}\\
    \cline{3-8}
    && CIFAR-10&CIFAR-100&Tiny ImageNet& CIFAR-10&CIFAR-100&Tiny ImageNet\\
    \midrule
    \rowcolor{gray!20}
    Uncompressed & 100\% & 94.74 & 78.70 & 66.14 & 93.54 & 74.77 & 62.65\\
    \midrule
    \multirow{3}{*}{Uniform} &50\%  &94.81 {\huge (-0.23)} &77.98 {\huge (-0.60)} &65.26 {\huge (-1.10)} &93.38 {\huge (-0.06)}&74.41 {\huge (-0.30)} &\textbf{61.87} {\huge (+0.08)}\\
    \cline{2-8}
    & 25\% &93.97 {\huge (-0.33)} &76.30 {\huge (-1.41)} &62.76 {\huge (-2.07)} &92.57 {\huge (-0.33)} &72.00 {\huge (-0.76)} &58.11 {\huge (-0.85)}\\
    \cline{2-8}
    & 12.5\% &92.71 {\huge (-0.55)} &73.89 {\huge (-0.96)} &60.34 {\huge (-1.50)} &90.98 {\huge (-0.46)} &67.51 {\huge (-2.22)} &51.90 {\huge (-2.16)}\\
    \midrule
    \multirow{3}{*}{Frobenius} & 50\%&94.71 {\huge (-0.35)} &78.28 {\huge (-0.30)} &65.98 {\huge (-0.38)} &93.40 {\huge (-0.04)} &74.58 {\huge (-0.13)} &61.33 {\huge (-0.46)}\\
    \cline{2-8}
    & 25\% & 94.23 {\huge (-0.07)}&76.38 {\huge (-1.33)} & 63.76 {\huge (-1.07)}&92.40 {\huge (-0.50)} &72.07 {\huge (-0.69)} &58.00 {\huge (-0.96)}\\
    \cline{2-8}
    & 12.5\% &92.65 {\huge (-0.61)} &74.32 {\huge (-0.53)} &61.14 {\huge (-0.70)} &90.32 {\huge (-1.12)} &68.02 {\huge (-1.71)} &51.92 {\huge (-2.14)}\\
    \midrule
    \multirow{3}{*}{\makecell{Loss-aware\\ (\method) }} &50\% &\textbf{95.04} &\textbf{78.58} & \textbf{66.36}&\textbf{93.44}  &\textbf{74.71} &61.79\\ 
    \cline{2-8} 
    & 25\% &\textbf{94.30} &\textbf{77.71} &\textbf{64.83} &\textbf{92.90} &\textbf{72.76} &\textbf{58.96}\\
    \cline{2-8}
    & 12.5\% & \textbf{93.26}&\textbf{74.85} &\textbf{61.84} &\textbf{91.44} &\textbf{69.73} & \textbf{54.06}\\
    \bottomrule
    \end{tabular}
    }
\end{table}

\section{Limitation and Future Work}
\label{sec:limit}

\begin{wraptable}[8]{r}[0em]{0.39\textwidth}
    \Huge
    \centering
    \resizebox{1.0\linewidth}{!}{
    \begin{tabular}{c|c|c|c}
    \toprule 
    Method (CIFAR-100)& Top-1 Acc. &Nonlinear latency& Total latency\\
    \midrule
    Original ResNet-18 &76.52 &12.64 s & 73.72 s\\
    \midrule
    \method~(b2) &76.93 &12.64 s & 45.76 s \\
    \midrule
    +SNL(-50\% ReLU) &76.72 &6.32 s & 39.44 s \\
    \midrule
    +SNL(-60\% ReLU) &76.27 &5.06 s & 38.18 s \\
    \bottomrule
    \end{tabular}
    }
    \caption{Extend~\method~with nonlinear layer optimization method SNL.}
    \label{tab:limit}
\end{wraptable}
\method~focuses on improving the HE computation efficiency, which accounts for 75\% total latency and is the bottleneck in the hybrid HE/MPC scheme. We can also extend~\method~with activation function optimization methods, e.g., ReLU pruning method SNL~\cite{cho2022SNL}. As shown in Table~\ref{tab:limit}, we prune $50\%$ ReLUs in~\method~(b2) without accuracy loss, achieving $2\times$ latency reduction in nonlinear layers. We regard a more in-depth study of joint linear/nonlinear layer optimization as our future work.

\section{Conclusion}
\label{sec:conclusion}

In this paper, we introduce \method, a network/protocol co-optimization framework to enhance the efficiency of HE-based DNN inference. \method~leverages block circulant transformation to reduce the HE computation. \method~features a novel encoding method,~\encode, and a latency-aware block size assignment algorithm. \method~significantly improves the network-level inference efficiency while maintaining a high accuracy. \method~achieves a latency reduction of $1.3\sim5.0 \times$ compared to Bolt in MobileNetV2, ResNet-18 and ViT. Moreover, when compared with SpENCNN, \method~attains up to $12\%$ higher accuracy, demonstrating a high potential to accelerate private inference across both ConvNets and Transformers.

\section{Acknowledgement}
This work was partly supported by Beijing Municipal Science and Technology Program (No. Z241100004224015), Ant Group, and the 111 Project (B18001).


\newpage
\bibliography{main}
\bibliographystyle{unsrt}
\newpage
\appendix
\onecolumn
\section{Related Works}\label{app:related_work}
To improve the efficiency of HE-based DNN inference, existing works mainly focus on optimizing the HE encoding algorithm~\cite{huang2022cheetah,hao2022iron,xu2023falcon,Juvekar_Vaikuntanathan_gazelle_2018,rathee2020cryptflow2,zhang2021gala,ju2023neujeans,pang2023bolt} and the DNN architectures~\cite{jha2021deepreduce,cho2022SNL,zeng2023copriv,kundu2023SENet,cho2022sphynx,peng2023autorep,zeng2022mpcvit,Cai_Zhang_Ning_Xin_Wu_Hunter,HE-PEx,spencnn,lou2020falcon}. 
HE encoding optimizations focus on improving the encoding density (i.e., useful elements per polynomial) to reduce communication~\cite{hao2022iron,xu2023falcon,lu2023bumblebee} and HE computations~\cite{huang2022cheetah,pang2023bolt,ju2023neujeans}. For example, Cheetah~\cite{huang2022cheetah} proposes an efficient rotation free encoding algorithm for convolutions and Falcon~\cite{xu2023falcon} further improve the communication efficiency for group-wise convolution. Iron~\cite{hao2022iron} and BubbleBee~\cite{lu2023bumblebee} optimize the encoding algorithm for general matrix multiplications (GEMMs). Neujeans~\cite{ju2023neujeans} and Bolt~\cite{pang2023bolt} further introduce the baby-step giant-step (BSGS) algorithm to reduce the number of HE rotations.

DNN architecture optimizations focus on developing HE-friendly architectures to improve inference efficiency including HE-friendly activation approximation or pruning \cite{jha2021deepreduce,cho2022SNL,zeng2023copriv,kundu2023SENet,cho2022sphynx,peng2023autorep}, weight pruning~\cite{Cai_Zhang_Ning_Xin_Wu_Hunter,HE-PEx,spencnn}, etc.~\cite{cho2022SNL,jha2021deepreduce,kundu2023SENet,cho2022sphynx,peng2023autorep} optimize the ReLU functions through pruning and approximation for communication and computation reduction.~\cite{zeng2022mpcvit,li2022mpcformer} propose to prune and approximate GeLU functions for efficient private transformer inference.
~\cite{spencnn,Cai_Zhang_Ning_Xin_Wu_Hunter,HE-PEx} propose HE-friendly structured pruning to reduce both HE rotations and multiplications.

\section{Baby-step Giant-step (BSGS) Algorithm for~\encode}\label{app:cirencode}
\begin{figure}[tbh]
    \centering
    \includegraphics[width=0.9\linewidth]{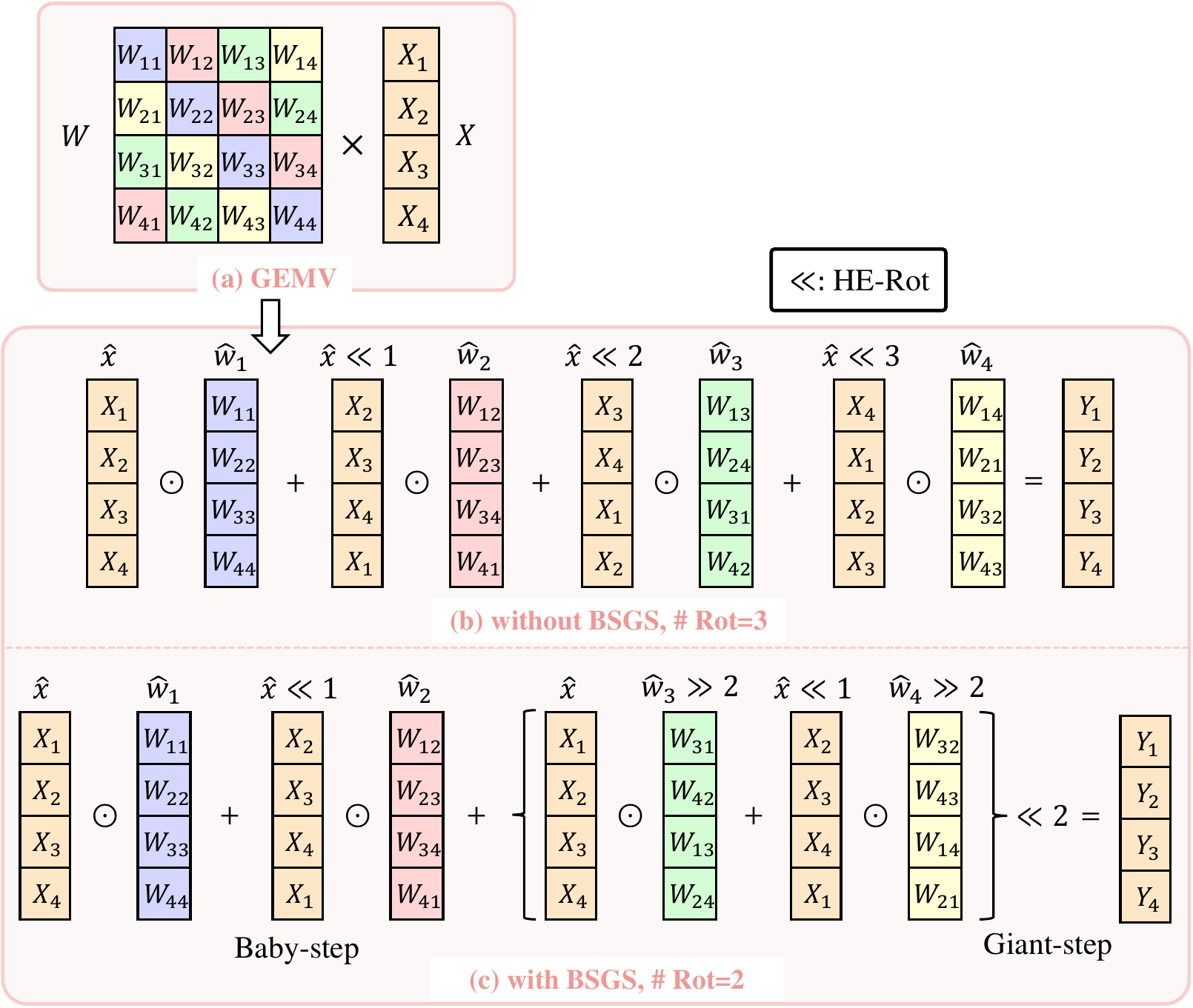}
    \caption{An example of GEMV using BSGS algorithm.}\label{fig:bsgs}
\end{figure}
\begin{figure}[!tb]
    \centering
    \includegraphics[width=0.5\linewidth]{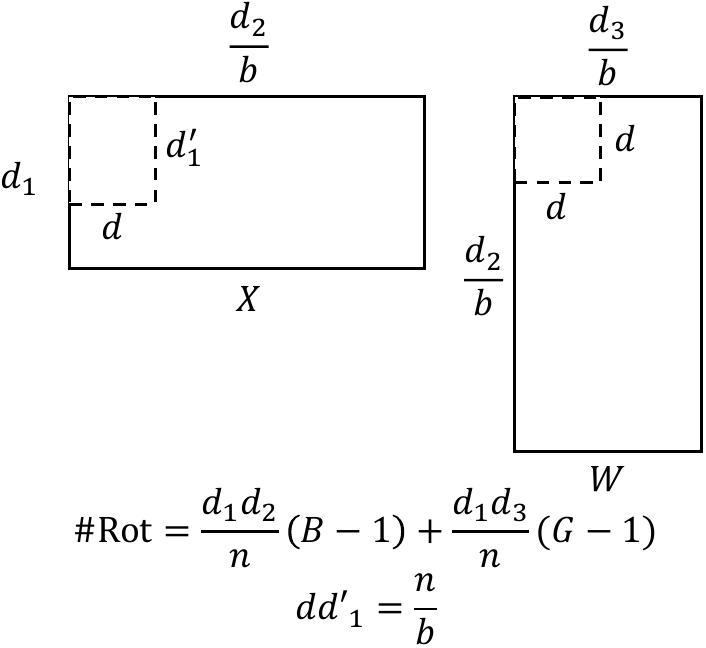}
    \caption{Illustration of our BSGS algorithm for block circulant GEMM with tiling.}\label{fig:bsgs2}
\end{figure}

The BSGS algorithm is used for GEMV and GEMM to reduce the number of HE rotations~\cite{pang2023bolt,ju2023neujeans}. We visualize the high-level idea of the BSGS algorithm in Figure~\ref{fig:bsgs}. Instead of rotating each input polynomial once, the BSGS algorithm divides the rotations into two steps: baby-step and giant-step which can be formulated as
\begin{equation}
    \sum_{j=1}^G\left(\sum_{i=1}^B {\hat{w}}_{(j-1) B+i}^{\text {diag }} \odot({\hat{x}}<<(i-1))\right)<<(j-1) B
\end{equation}
Here, $G,B$ are the number of giant-step and baby-step, respectively. The total number of rotations is reduced to $B+G-2$. In GEMM with dimension $(d_1,d_2,d_3)$, tiling is needed to split matrices into smaller blocks whose maximum size is HE polynomial degree $n$. Moreover, when extend BSGS to~\encode, the dimension of GEMM becomes $(d_1,\frac{d_2}{b},\frac{d_3}{b})$ and the polynomial degree becomes $\frac{n}{b}$. We do not encode the $d_1$ dimension into each circulant block, instead, we treat the computation cross blocks as a GEMM and use the BSGS algorithm to determine the tiling size of the $d_1$ dimension. Therefore, how to tile and choose $B,G$ is crucial to minimize the number of rotations. We propose to formulate this optimization problem as a nonlinear programming problem as
\begin{equation}\label{eq:bsgs}
    \begin{aligned}
        \min \quad&  \text{\# Rot}=\frac{d_1d_2}{n}(B-1)+\frac{d_1d_3}{n}(G-1)\\
         \text{s.t. }\quad& B*G=d \\
        & d_1'd= \frac{n}{b} \\
        & d_1'\le d_1 \\
        & d\le \min(\frac{d_3}{b},\frac{d_2}{b}) 
    \end{aligned}
\end{equation}
We give an illustration of our BSGS algorithm in Figure~\ref{fig:bsgs2}. The tile sizes of input and weight are $(d_1^\prime,d)$ and $(d,d)$, respectively. The constraints in Equation~\ref{eq:bsgs} are derived from a tile containing at most $n$ elements and a tile size cannot exceed the size of the matrix.
This problem has a small solution space. With $B,G\le \min(\frac{d_3}{b},\frac{d_2}{b})$, The solution space is at most $\min(\frac{d_3}{b},\frac{d_2}{b})^2$, allowing us to
directly solve it using a search algorithm with the complexity
of $O((\min(\frac{d_3}{b},\frac{d_2}{b})^2))$. Our experiments show that the search algorithm can find the optimal solution within milliseconds for all cases. 

\textbf{Complexity analysis of \# Rot.} We proof in Equation~\ref{eq:bsgs_complexity} that the complexity of \#Rot with our BSGS algorithm is $O(\sqrt{d_1d_2d_3/(nb)})$. 
\begin{equation}\label{eq:bsgs_complexity}
    \begin{aligned}
        \text{\# Rot}&=\frac{d_1d_2}{n}(B-1)+\frac{d_1d_3}{n}(G-1)\\
        &\ge 2\frac{d_1}{n}\sqrt{d_2d_3(B-1)(G-1)}\\
        &\iff d_2(B-1)=d_3(G-1)\\
        & O(\text{\# Rot})=O(\frac{d_1}{n}\sqrt{d_2d_3d})\\
        &=O(\frac{d_1}{n}\sqrt{d_2d_3n/bd_1}) \\
        &=O(\sqrt{d_1d_2d_3/(nb)}) \\
    \end{aligned}
\end{equation}
Here we omit the last constraint in Equation~\ref{eq:bsgs} for simplicity.  

\textbf{Complexity analysis of \# Mul.} The complexity of \# Mul is given by Equation~\ref{eq:mul_complexity_gemm}.
\begin{equation}\label{eq:mul_complexity_gemm}
    \begin{aligned}
        & O(\text{\# Mul})=O(\frac{d_2}{b}\cdot \frac{d_3}{b}\cdot \frac{bd_1}{n})\\
        &=O(d_1d_2d_3/(nb)) \\
    \end{aligned}
\end{equation}

\textbf{Boundary cases.} When $d_1\min(\frac{d_3}{b},\frac{d_2}{b})<\frac{n}{b}$, the tile size of input will be $d_1\min(\frac{d_3}{b},\frac{d_2}{b})$ although it's not often the case. In addition, the second constraint in Equation~\ref{eq:bsgs} should actually be $\left[ d_1^\prime b \right]_{2^k}d=n$. $\left[ \cdot \right]_{2^k}$ means the next nearest power of 2. This is because NTT requires the input size to be a power of 2. Consequently, we consider all these boundary conditions in the search algorithm in practice.

\section{\encode~for Convolutions}\label{app:encode_conv}
In this section, we extend~\encode~to convolutions. We denote the input, weight and output of a block circulant convolution operation as $X\in \mathbb{Z}^{C\times H\times W}, W\in \mathbb{Z}^{K\times C\times R\times R}, Y=W\circledast X\in \mathbb{Z}^{K\times (H-R+1)\times (W-R+1)}$. Here $\circledast$ represents the convolution operation. We assume stride=1 for simplicity. $W$ is a block circulant matrix with respect to the first two dimensions with block size $b$.

\begin{figure}[!tb]
    \centering
    \includegraphics[width=1.0\linewidth]{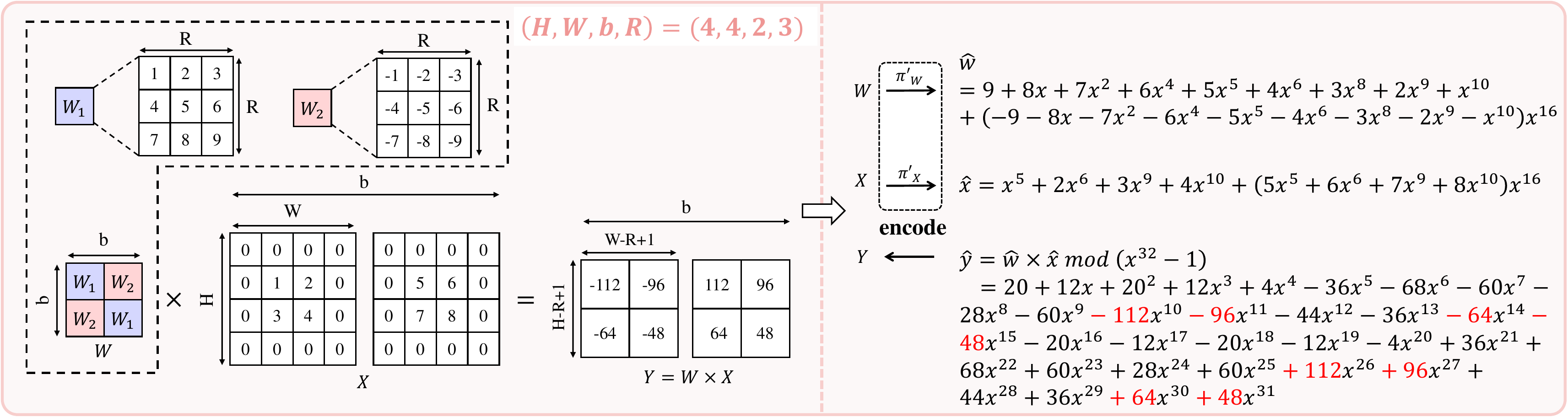}
    \caption{A toy example of CirEncode within a circulant convolution where $(H,W,b,R) = (4,4,2,3)$.}\label{fig:cirencode_conv}
\end{figure}

\textbf{Encoding within a circulant block.} For each circulant block, we define two encoding functions $\pi'_\mathrm{X}: \mathbb{Z}^{b\times H\times W}\rightarrow \mathbb{A}_n$ and $\pi'_\mathrm{W}: \mathbb{Z}^{b\times b\times R\times R}\rightarrow \mathbb{A}_n$ as follows:
\begin{align*}
    & \hat{x}=\pi'_{\mathrm{X}}({X}) \quad \mathrm{s.t.} \quad \hat{x}[iHW+jW+k]=X[i,j,k], i\in [b], j\in [H], k\in [W]\\
    & \hat{w}=\pi'_{\mathrm{W}}({W}) \quad \mathrm{s.t.}\quad \hat{w}[iHW+(W+1)(R-1)-jW-k]=W[i,0,j,k], i\in [b], j\in [R], k\in [R]
\end{align*}
where other coefficients of $\hat{w}$ are set to 0. Multiplication of polynomials $\hat{y}=\hat{w}\times \hat{x}$ directly gives the result of ${Y}={W\circledast X}$ as described in Theorem~\ref{theorem:encode_conv}. We defer the proof to Appendix~\ref{proof:encode_conv}.
\begin{theorem}\label{theorem:encode_conv}
    Assuming $HWb\le n$, given a circulant convolution kernel $W\in \mathbb{Z}^{b\times b\times R\times R}$ and input tensor $X\in \mathbb{Z}^{b\times H\times W}$. Define two polynomials $\hat{w}=\pi'_{\mathrm{W}}({W})$ and $\hat{x}=\pi'_{\mathrm{X}}({X})$. The polynomial multiplication result $\hat{y}=\hat{w}\times \hat{x}$ directly maps to the result of ${Y}={W\circledast X}\in \mathbb{Z}^{b\times (H-R+1)\times (W-R+1)}$ where ${Y}[i,j,k]=\hat{y}[iHW+(W+1)(R-1)+jW+k]$.
\end{theorem}
We show a toy example of~\encode~for circulant convolution in Figure~\ref{fig:cirencode_conv}.

\textbf{Encoding across circulant blocks.} Consider each circulant block with input dimension $(b,H,W)$ and weight dimension $(b,b,R,R)$ as a basic unit. The computation across circulant blocks can be regarded as a GEMV with dimension $(1,\frac{C}{b},\frac{K}{b})$. Then we leverage SIMD diagonal encoding which is the same as the block circulant GEMM.

\begin{figure}[!tb]
    \centering
    \includegraphics[width=0.4\linewidth]{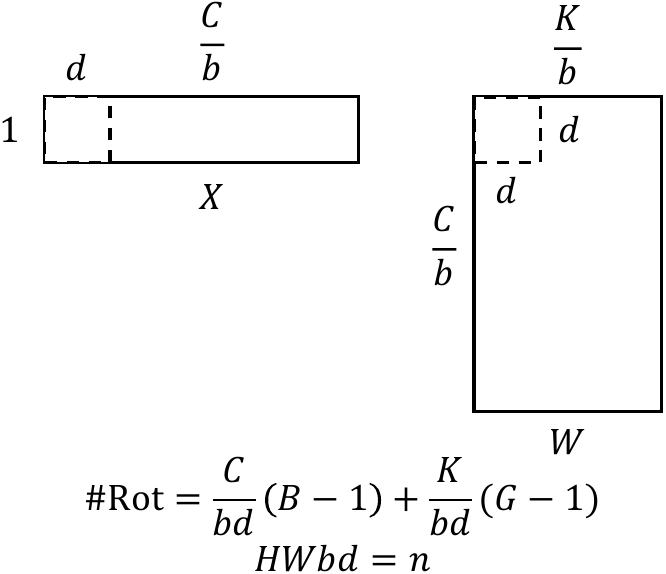}
    \caption{Illustration of our BSGS algorithm for block circulant convolution with tiling.}\label{fig:bsgs3}
\end{figure}

\textbf{BSGS algorithm for block circulant convolution.} Similar to block circulant matrix multiplication, the BSGS algorithm for block circulant convolution can be formulated as an non-linear programming problem as
\begin{equation}\label{eq:bsgs_conv}
    \begin{aligned}
        \min \quad&  \text{\# Rot}=\frac{HWC}{n}(B-1)+\frac{HWK}{n}(G-1) \\
         \text{s.t. }\quad& B*G=d \\
         & HWbd=n \\
         & d\le \min(\frac{C}{b},\frac{K}{b}) \\
        \end{aligned}
\end{equation}
We give an illustration in Figure~\ref{fig:bsgs3} where the tile sizes of input and weight are $(1,d)$ and $(d,d)$, respectively. This problem has a small solution space. With $B,G\le \min(\frac{C}{b},\frac{K}{b})$, The solution space is at most $(\min(\frac{C}{b},\frac{K}{b}))^2$, allowing us to
directly solve it using a search algorithm with a complexity
of $O((\min(\frac{C}{b},\frac{K}{b}))^2)$. Our experiments show that the search algorithm can find the optimal solution within milliseconds for all cases. 

\textbf{Complexity analysis of \# Rot.} We proof in Equation~\ref{eq:bsgs_complexity_conv} that the complexity of \# Rot in block circulant convolution with our BSGS algorithm is $O(\sqrt{HWCK/(nb)})$. 
\begin{equation}\label{eq:bsgs_complexity_conv}
    \begin{aligned}
        \text{\# Rot}&=\frac{HWC}{n}(B-1)+\frac{HWK}{n}(G-1) \\
        &\ge 2\frac{HW}{n}\sqrt{CK(B-1)(G-1)}\\
        &\iff C(B-1)=K(G-1) \\
        & O(\text{\# Rot})=O(\frac{HW}{n}\sqrt{CKd})\\
        &=O(\frac{HW}{n}\sqrt{\frac{CKn}{HWb}}) \\
        &=O(\sqrt{\frac{HWCK}{nb}}) \\
    \end{aligned}
\end{equation}
Here we omit the last constraint in Equation~\ref{eq:bsgs_conv} for simplicity. 

\textbf{Complexity analysis of \# Mul.} The complexity of \# Mul is given by Equation~\ref{eq:mul_complexity_conv}.
\begin{equation}\label{eq:mul_complexity_conv}
    \begin{aligned}
        & O(\text{\# Mul})=O(\frac{C}{b}\cdot \frac{K}{b}\cdot \frac{HWb}{n})\\
        &=O(HWCK/(nb)) \\
    \end{aligned}
\end{equation}

\section{Why does structured pruning fail in BSGS algorithm?}\label{app:spencnn_fail}
\begin{figure}[!tb]
    \centering
    \includegraphics[width=0.8\linewidth]{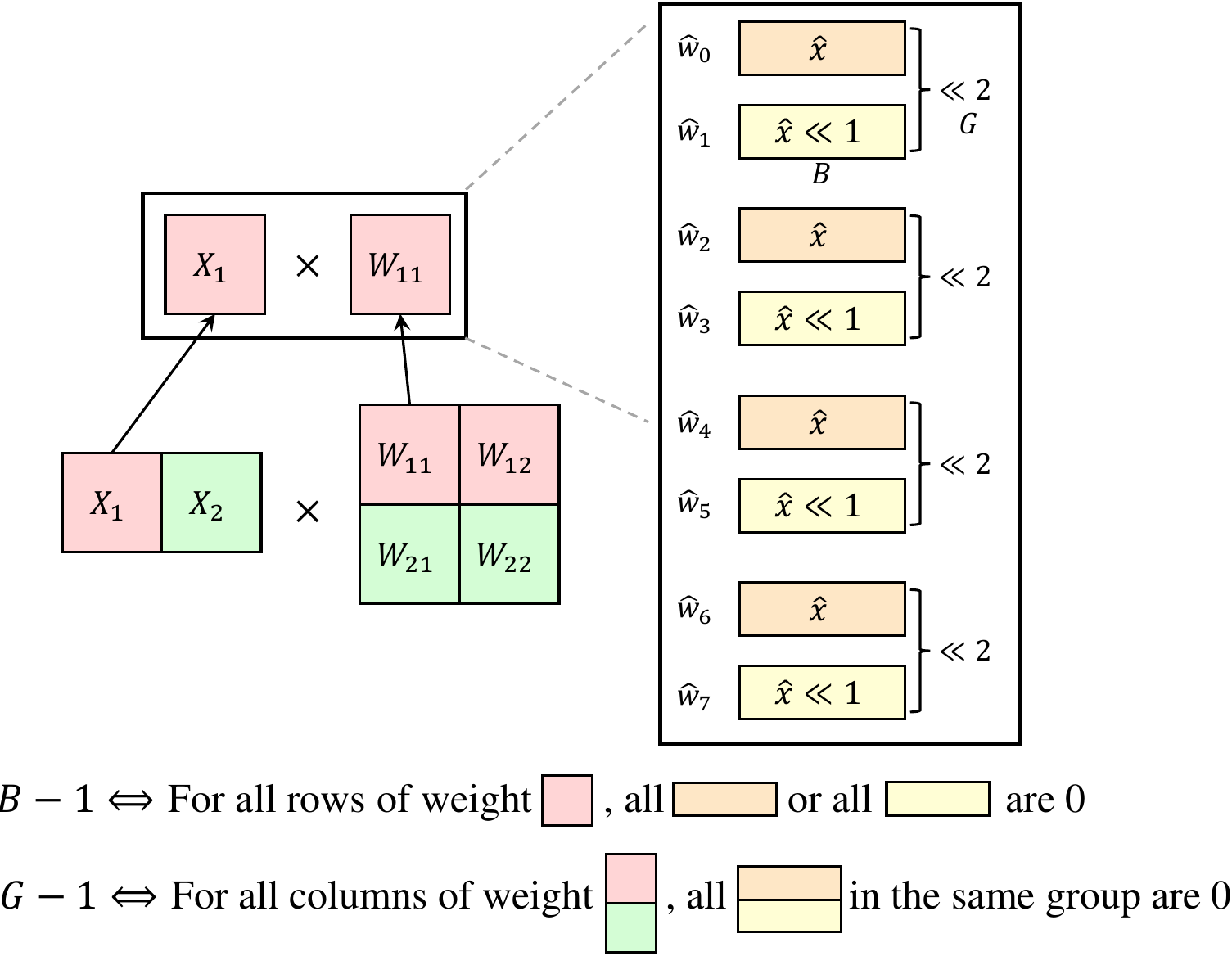}
    \caption{Illustration of the limitation of structured pruning in BSGS algorithm.}\label{fig:spencnn_fail}
\end{figure}
HE-friendly structured pruning~\cite{Cai_Zhang_Ning_Xin_Wu_Hunter,spencnn} reduces the number of rotations by pruning the diagonals of weight matrices. However, this technique is not feasible in the BSGS algorithm. Figure~\ref{fig:spencnn_fail} demonstrates the limitations of structured pruning in BSGS. To illustrate, consider a GEMM where input and weight matrices are tiled into smaller blocks, such as $X_1, X_2$ and $W_{11}, W_{12}, W_{21}, W_{22}$. First focusing on the multiplication between $X_1$ and $W_{11}$, note that in BSGS, rotations are split into baby-step and giant-step. Assuming $B=2, G=4$, there are four groups, each containing two ciphertexts $(\hat{x}, \hat{x}\ll 1)$, and eight weight polynomials $\hat{w}_0,\ldots,\hat{w}_7$ which are the eight diagonals of the weight matrix $W_{11}$. Each group requires one baby-step rotation to achieve $\hat{x} \ll 1$ and one giant-step rotation. Pruning diagonals to reduce rotations in BSGS is challenging. For instance, to reduce a baby-step rotation, all diagonals in the same position across different groups, such as $\hat{w}_1, \hat{w}_3, \hat{w}_5, \hat{w}_7$, must be pruned. Additionally, considering tiling, $X_1$ must multiply with all weight matrices in the first row, i.e., $W_{11}, W_{12}$. 
Thus, to decrease a single baby-step rotation, diagonals in the same position across all groups for all first-row weight matrices must be pruned. A similar challenge exists for giant-step rotations; to reduce one giant-step rotation, entire groups like $\hat{w}_0,\hat{w}_1$, in all first-column of the weight matrices must be pruned. Consequently, it is difficult for existing structured pruning methods to meet these constraints, leading to the limitation of reducing the number of rotations.

\section{An example of our loss-aware initialization for circulant matrices}\label{app:initialization}
We give an example of our circulant transformation initialization in Equation~\ref{eq:initialization}. The input matrix $W$ is a $2\times 2$ matrix and the values of $W$ and $\frac{\partial \mathcal{L}(\mathcal{D})}{\partial W}$ are artificial for simplicity.
\begin{equation}\label{eq:initialization}
    \begin{aligned}
        W=\left[ \begin{array}{cc}
            1 & 2 \\
            4 & 3 \\
            \end{array} 
            \right ], \left(\frac{\partial \mathcal{L}(\mathcal{D})}{\partial W} \right)&=\left[ \begin{array}{cc}
                1 & 2  \\
                3 & 5  \\
                \end{array} 
                \right ] \\
        \min \left \|W^\prime-W\right\|^2_2 \Rightarrow W^\prime &= \left[ \begin{array}{ccc}
            2 & 3 \\
            3 & 2 \\
            \end{array} 
            \right ] \\
        \min \Omega_i \Rightarrow W^\prime=\mathbb{E}\left[ \begin{array}{cc}
            1*1^2 & 2*2^2 \\
            4*3^2 & 3*5^2 \\
            \end{array} 
            \right ]_{diag} &= \left[ \begin{array}{cc}
                \frac{1*1^2+3*5^2}{1^2+5^2} & \frac{2*2^2+4*3^2}{2^2+3^2} \\
                \frac{2*2^2+4*3^2}{2^2+3^2} & \frac{1*1^2+3*5^2}{1^2+5^2} \\
                \end{array} 
                \right ]=\left[ \begin{array}{cc}
                    2.92 & 3.38 \\
                    3.38 & 2.92 \\
                    \end{array} 
                    \right ]
                 \\
    \end{aligned}
\end{equation}

\section{Inverted Residual Fusion Algorithm}\label{app:inverted_residual_fusion}
The key idea of the inverted residual fusion is to compute consecutive linear layers at once with one round communication. The algorithm is described in Algorithm~\ref{alg:inverted_residual_fusion} where $\langle \cdot \rangle^C, \langle \cdot \rangle^S$ are the secret shares held by the client and the server. $\boxplus, \boxminus, \boxtimes$ represent homomorphic addition, subtraction, and multiplication, respectively. 

\begin{algorithm}[h]
\DontPrintSemicolon
    \SetAlgoLined
    \KwIn{Client holds $ \langle \bm{X}_1 \rangle^C$, and Server holds $\langle \bm{X}_1 \rangle^S,\Enc(\bm{X}_{res}), \bm{W}_1$ and $\bm{W}_2$.}
    \KwOut{Client and Server get $\langle \bm{Y}_2 \rangle^C,\langle \bm{Y}_2 \rangle^S$, respectively, where $\bm{Y}_2=\text{ConvBN}(\bm{W}_2, \bm{X}_{res}+\text{ConvBN}(\bm{W}_1,\bm{X}_1))$.}

    Client encodes and encrypts $\langle \bm{X}_1 \rangle^C$ as $\Enc(\langle \bm{X}_1 \rangle^C)$ and sends it to Server.\;

    Server computes $\Enc(\bm{Y}_{1})=\bm{W}_1\boxtimes[\Enc(\langle \bm{X}_1 \rangle^C)\boxplus \langle \bm{X}_1 \rangle^S]$. \;

    Server computes $\Enc(\bm{X}_{res}+\bm{Y}_{1})=\Enc(\bm{X}_{res})\boxplus\Enc(\bm{Y}_{1})$.\;

    Server computes $\Enc(\bm{Y}_{2})=\bm{W}_2 \boxtimes \Enc(\bm{X}_{res}+\bm{Y}_{1})$.\;

    Server samples random noise $\bm{R}$ which has the same shape as $\bm{Y}_{2}$. Server then computes $\Enc(\bm{Y}_{2}-\bm{R})=\Enc(\bm{Y}_{2})\boxminus \bm{R}$.\;

    Server sends $\Enc(\bm{Y}_{2}-\bm{R})$ to Client and sets $\langle \bm{Y}_{2} \rangle^S=\bm{R}$.\;

    Client decrypts $\Enc(\bm{Y}_{2}-\bm{R})$ to get $\langle \bm{Y}_{2} \rangle^C=\bm{Y}_{2}-\bm{R}$.\;
    
    \caption{Inverted Residual Fusion Algorithm}
    \label{alg:inverted_residual_fusion}
\end{algorithm}









    

\section{Details of Experimental Setup}\label{app:exp_details}
\subsection{Network Architectures}\label{app:network_architectures}
We evaluate~\method~on MobileNetV2~\cite{sandler2018mobilenetv2}, ResNet-18~\cite{he2016resnet}, and ViT~\cite{hassani2021escaping}. The detailed architectures across different datasets are in Table~\ref{tab:benchmarks}. It should be noted that for ViT, we use ViT-lite architectures from~\cite{hassani2021escaping}.
\begin{table}[!tb]
    \caption{\method~evaluation benchmarks.}\label{tab:benchmarks}
    \centering
    \resizebox{1.0\linewidth}{!}{
    \begin{tabular}{cccccc}
        \toprule
        Model  &  Layers &  \# Params (M) & MACs (G) & Dataset \\
        \midrule
        MobileNetV2 & 52 CONV, 1 FC, 1 AP, 35 ReLU  & 2.24 & 0.093 & CIFAR/Tiny ImageNet \\
        MobileNetV2 &  52 CONV, 1 FC, 1 AP, 35 ReLU &  3.5 & 0.32 & ImageNet \\
        ResNet-18 & 52 CONV, 1 FC, 1 AP, 35 ReLU & 11.17 & 0.558 & CIFAR/Tiny ImageNet \\
        ViT &  Hidden Dim=256, Number of blocks=7&  3.72 & 0.24 & CIFAR \\
        ViT &  Hidden Dim=192, Number of blocks=9&  2.77 & 0.69 & Tiny ImageNet \\
        \bottomrule
    \end{tabular}
    }
\end{table}
\subsection{Training Details}\label{app:training_details}
All baseline methods and~\method~are trained using identical hyper-parameters, including data augmentation, number of epochs, and others. These hyper-parameters are detailed in the `configs' folder within our codebase.
We also leverage self knowledge distillation to guide the training of the circulant networks and the pruned networks. 

\subsection{Computational Resources in Experiments}
For CIFAR and Tiny ImageNet datasets, we train all models on a single NVIDIA RTX4090 GPU and a single NVIDIA A6000 GPU. For ImageNet, we train all models on 8 NVIDIA A100 GPUs. The epochs are 300 and the total training time is around 1 day for CIFAR and Tiny ImageNet as well as ImageNet datasets.

\section{Additional Experimental Results}\label{app:add_exp}

\subsection{Latency breakdown of~\method}\label{app:exp_breakdown}
\begin{figure}[!tb]
    \centering
    \includegraphics[width=0.7\textwidth]{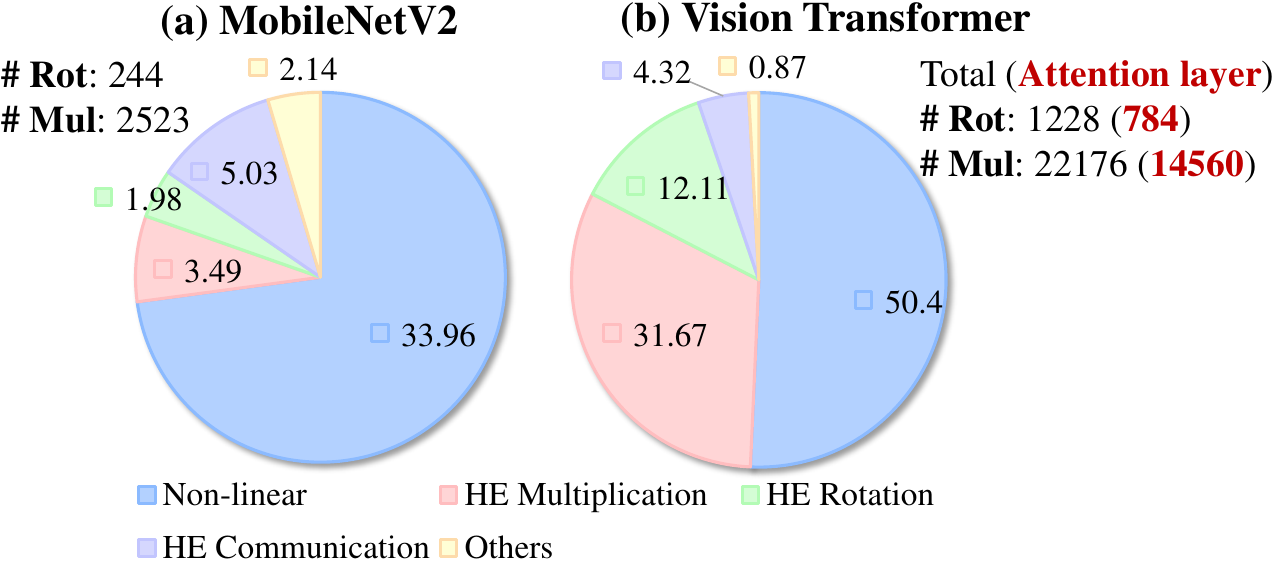}
    \caption{Latency (s) breakdown of~\method~(b8) on MobileNetV2 and ViT on CIFAR-10.}\label{fig:exp_breakdown}
\end{figure}
In Figure~\ref{fig:exp_breakdown}, we present the latency breakdown of~\method~(b8) applied to MobileNetV2 and ViT on CIFAR-10. It is observed that~\method significantly reduces the latency associated with HE rotations and multiplications, shifting the bottleneck to nonlinear layers. Furthermore, in ViT, the self-attention layers account for a large proportion of the total HE operations. Since these layers lack weight matrices, they cannot benefit from block circulant transformations.

\subsection{Results on more networks}
\begin{figure}[!tb]
    \centering
    \includegraphics[width=0.9\linewidth]{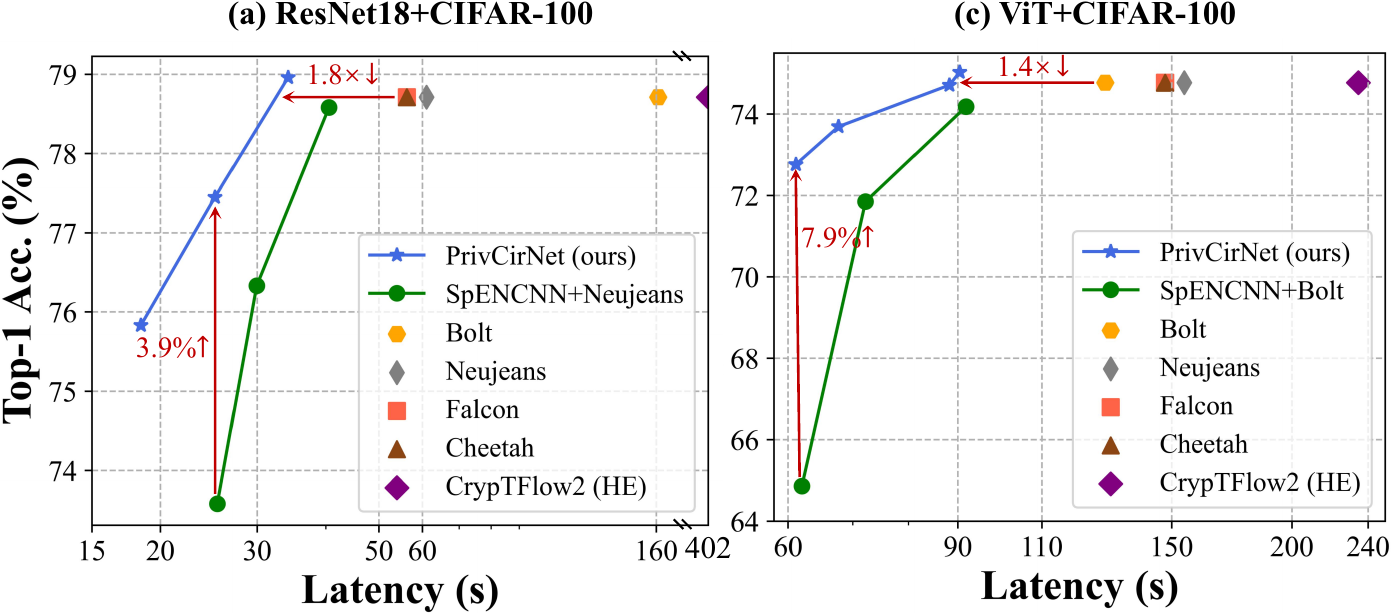}
    \caption{Comparison with SpENCNN and other prior-art protocols on ResNet-18 and ViT on CIFAR-100.} 
    \label{fig:app_c100}
\end{figure}

\begin{figure}[!tb]
    \centering
    \includegraphics[width=1.0\linewidth]{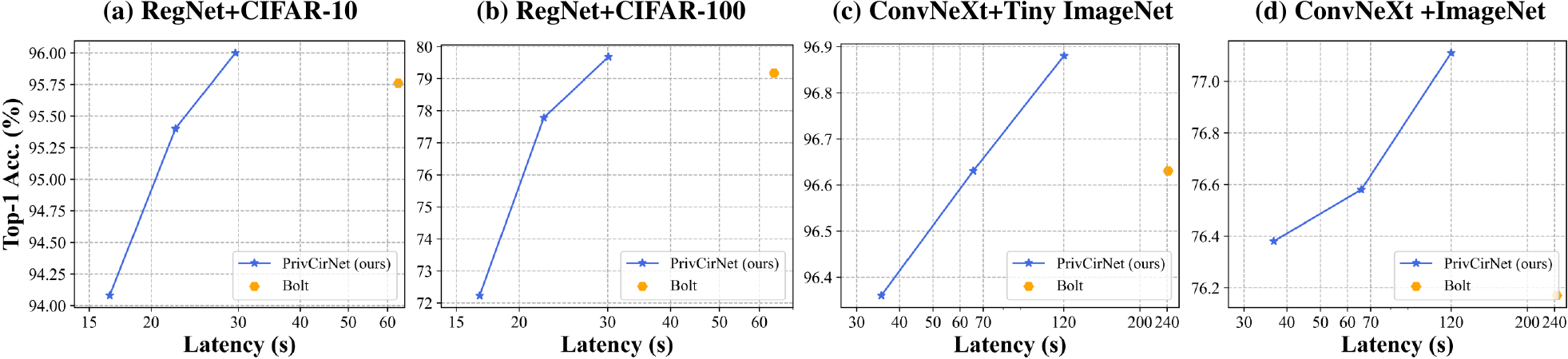}
    \caption{Comparison with Bolt on RegNet and ConvNeXt.} 
    \label{fig:regnet}
\end{figure}

\begin{figure}[!tb]
    \centering
    \includegraphics[width=0.9\linewidth]{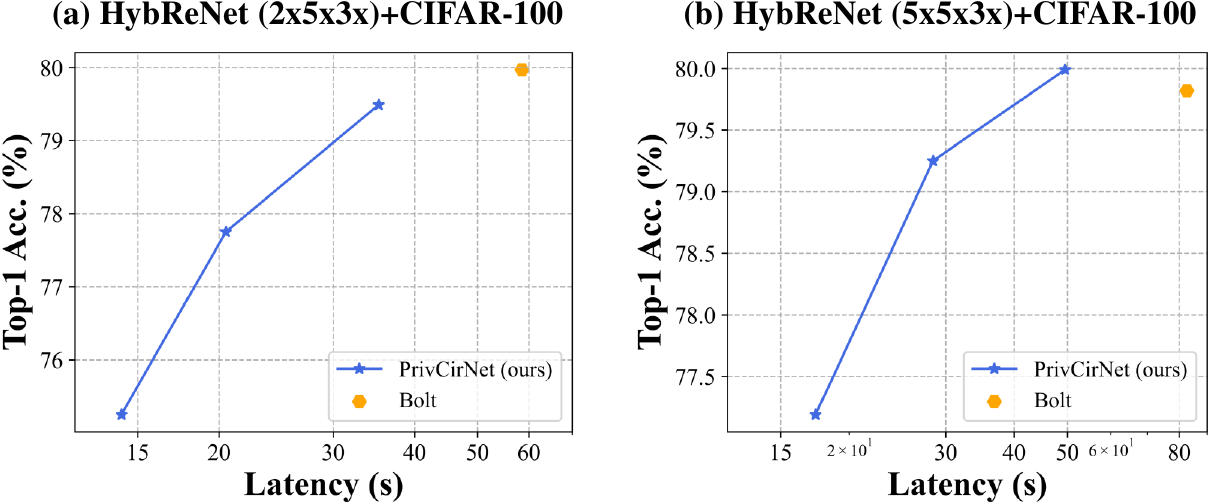}
    \caption{Result of applying~\method on DeepReshape.} 
    \label{fig:deepreshape}
\end{figure}
\begin{table}[t]
    \centering
    \Huge
    \caption{Accuracy and latency results when combining PrivCirNet and DeepReshape (ReLU reduction).}
    \label{tab:deepreshape}
    \resizebox{1.0\linewidth}{!}{
        \begin{tabular}{c|c|c|c}
        \toprule
        \multirow{2}{*}{Method} & \multirow{2}{*}{Top-1 Acc.} & Linear layers' latency  & Nonlinear layers' latency \\
        & &  (s)& (s) \\
        \midrule 
        HybReNet (5x5x3x) + PrivCirNet (b2) & 79.99 & 49.52 & 9.4 \\
        +DeepReshape (-53\% ReLU) & 79.67 & 49.52 & 4.4 \\
        \midrule 
        HybReNet (2x5x3x) + PrivCirNet (b2) & 79.49 & 35.25 & 9.0 \\
        +DeepReshape (-50\% ReLU) & 79.03 & 35.25 & 4.5 \\
        +DeepReshape (-72\% ReLU) & 77.91 & 35.25 & 2.5 \\
        \bottomrule 
        \end{tabular}

    }
\end{table}

\paragraph{Results of ResNet-18 and ViT on CIFAR-100}
In Figure~\ref{fig:app_c100}, we show the results of ResNet-18 and ViT on CIFAR-100. We compare~\method~with SOTA HE-based DNN inference frameworks and HE-friendly structured pruning method SpENCNN. We find that: \textbf{\underline{1)}}~\method~achieves $1.8\times$ latency reduction on ResNet-18 and $1.4\times$ latency reduction on ViT compared with SOTA frameworks Cheetah and Bolt with iso-accuracy. \textbf{\underline{2)}} Compared with SpENCNN,~\method~achieves $3.9\%$ and $7.9\%$ higher accuracy on ResNet-18 and ViT with lower latency, respectively. \textbf{\underline{3)}} Bolt performs worse than Cheetah in ResNet-18 because Bolt needs to transform convolution into GEMM which increases the hidden dimension by $9\times$ in $3\times 3$ convolutions. By contrast,~\method~support both convolution and GEMM efficiently.

\paragraph{Results of RegNet and ConvNeXt} 
In Figure~\ref{fig:regnet}, we show the results of RegNet~\cite{radosavovic2020regnet} and ConvNeXt~\cite{liu2022convnet} on CIFAR. We compare~\method~with SOTA HE-based DNN inference framework Bolt.

\paragraph{Comparion with DeepReshape~\cite{jha2023deepreshape}} 

DeepReshape optimizes ReLUs and FLOPs by designing a series of more FLOPs-efficient networks, dubbed HybReNets while pruning the ReLU layers. DeepReshape achieves a better latency-ReLU trade-off than SENet~\cite{kundu2023SENet}, SNL~\cite{cho2022SNL}, etc. DeepReshape and PrivCirNet are orthogonal and can be applied together to further reduce the inference latency.

In Figure~\ref{fig:deepreshape} and Table~\ref{tab:deepreshape}, we show the application of PrivCirNet to HybReNets on CIFAR-100, which also yields promising results. We apply the ReLU pruning method proposed in DeepReshape to reduce the latency of nonlinear layers. 

From the results, we can see that PrivCirNet is effective when combined with DeepReshape, achieving significant latency reduction in both linear and nonlinear layers. 

\paragraph{Discussion on the impact of selecting different baseline networks} 
The varying accuracy degradation observed across different baseline networks (MobileNetV2, ResNet, HybReNet, RegNet, ConvNeXt) can be partly attributed to the differing proportions of parameters occupied by standard convolutional layers. For instance, in ConvNeXt, 98\% of the parameters are derived from standard convolution, with less than 2\% from depth-wise/group-wise convolution, providing significant compression potential using PrivCirNet. In contrast, standard convolution parameters account for only 64\% and 78\% of RegNet and MobileNetV2, respectively. As a result, RegNet and MobileNetV2 exhibit larger accuracy degradation at higher compression rates. 
\section{Proofs}

\subsection{Proof of Theorem~\ref{theorem:encode}}\label{proof:encode}
For a given input matrix $X$ and a circulant matrix $W$, we have
\begin{equation}
    \begin{aligned}
        & W\in \mathbb{Z}^{b\times b}, W[i,j]=W[0,(b-i+j)\mod b], \forall i\in [b], \forall j\in[b] \\
        & X\in \mathbb{Z}^{b\times d_1}, X[i,j], \forall i\in [b], \forall j\in [d_1] \\
    \end{aligned}
\end{equation}
The matrix multiplication result $Y$ is
\begin{equation}
    \begin{aligned}
        Y=WX\in \mathbb{Z}^{b\times d_1}, Y[i,j]=\sum_{k=0}^{b-1}W[i,k]X[k,j]=\sum_{k=0}^{b-1}W[0,(b-i+k)\mod b]X[k,j]
    \end{aligned}
\end{equation}
The polynomials $\hat{x}=\pi_{\mathrm{X}}({X}), \hat{w}=\pi_{\mathrm{W}}({W})$ after~\encode~are
\begin{equation}
    \begin{aligned}
        & \hat{x}\in \mathbb{A}_n, \hat{x}[id_1+j]=X[i,j], \forall i\in[b], \forall j\in [d_1]\\
        & \hat{w}\in \mathbb{A}_n, \hat{w}[id_1]=W[i,0]=W[0,(b-i)\mod b], \forall i\in [b],\\
    \end{aligned}
\end{equation}
The other slots of $\hat{w}$ are set to 0. The polynomial multiplication result $\hat{y}=\hat{w}\times \hat{x}$ directly gives the matrix multiplication result $Y$ as
\begin{equation}
    \begin{aligned}
        \hat{y}&=\hat{w}\times \hat{x}\in \mathbb{A}_n\\
        \hat{y}[id_1+j]&=\sum_{k=0}^{b-1}\hat{w}[(i-k)d_1]\hat{x}[kd_1+j]\\
        &=\sum_{k=0}^{b-1}W[0,(b-i+k)\mod b]X[k,j]\\
        &=\sum_{k=0}^{b-1}W[i,k]X[k,j]=Y[i,j]\\
    \end{aligned}
\end{equation}
Besides, we extend the definition of $\hat{w}[i]=\hat{w}[bd_1+i],\forall i<0$.

\paragraph{Explanation of CirEncode Modulo $x^n - 1$}.~\encode~performs Discrete Fourier Transform (DFT) modulo $x^n-1$ on the plaintext. After the DFT, it applies SIMD encoding to enable element-wise multiplication. The correctness is demonstrated by the equation $\operatorname{DFT}(\hat{w})\odot \operatorname{DFT}(\hat{x})=\operatorname{DFT}(\hat{w}\times \hat{x} \mod x^n-1)$ 

\subsection{Proof of Theorem~\ref{theorem:encode_across}}\label{proof:encode_across}
Given $M$ circulant weight matrices $W_0, \ldots, W_{M-1} \in \mathbb{Z}^{b\times b}$ and input matrices $X_0, \ldots, X_{M-1} \in \mathbb{Z}^{b\times d_1}$, define the polynomials $\hat{w}_m=\pi_{\mathrm{W}}({W_m})$ and $\hat{x}_m=\pi_{\mathrm{X}}({X_m})$ with $m \in [M]$ following the coefficient packing in Theorem~\ref{theorem:encode}. We have:
\begin{equation}\label{eq:proof_across}
\begin{aligned}
    &\lj \operatorname{DFT}(\hat{w}_0)|\ldots|\operatorname{DFT}(\hat{w}_{M-1})\rj_{\mathrm{SIMD}}\times \lj \operatorname{DFT}(\hat{x}_0)|\ldots|\operatorname{DFT}(\hat{x}_{M-1})\rj_{\mathrm{SIMD}}\\
    &=\lj \operatorname{DFT}(\hat{w}_0)\odot \operatorname{DFT}(\hat{x}_0)| \ldots | \operatorname{DFT}(\hat{w}_{M-1})\odot \operatorname{DFT}(\hat{x}_{M-1})\rj_{\mathrm{SIMD}}\\
    &=\lj \operatorname{DFT}(\hat{w}_0\times \hat{x}_0)|\ldots |\operatorname{DFT}(\hat{w}_{M-1}\times \hat{x}_{M-1})\rj_{\mathrm{Coeff}} \\
    &=\lj\operatorname{DFT}(\hat{y}_0)|\ldots |\operatorname{DFT}(\hat{y}_{M-1})\rj_{\mathrm{Coeff}} \\
\end{aligned}
\end{equation}
Then we can perform inverse DFT and directly extract elements following Theorem~\ref{theorem:encode} from $\hat{y}_m$ to get $Y_m$, $\forall m \in [M]$. The second and the third lines of Equation~\ref{eq:proof_across} follow directly from Lemma~\ref{lemma:DFT} while the last line is derived from Theorem~\ref{theorem:encode}. Through Equation~\ref{eq:proof_across}, we simultaneously evaluate $M$ circulant GEMMs with \encode.

\subsection{Proof of Theorem~\ref{theorem:encode_conv}}\label{proof:encode_conv}

For a given input $X$ and a circulant weight $W$ of a convolution, we have
\begin{equation}
    \begin{aligned}
        W\in \mathbb{Z}^{b\times b\times R\times R}, W[i,j,:,:]&=W[0,(b-i+j)\mod b,:,:] \\
        &=W[(b-j+i)\mod b,0,:,:], \forall i \in [b], \forall j\in[b] \\
        X\in \mathbb{Z}^{b\times H\times W}, X[i,j,k], &\forall i \in [b], \forall j\in [H], \forall k\in [W]  \\
    \end{aligned}
\end{equation}
The convolution result $Y$ is
\begin{equation}
    \begin{aligned}
        &Y=W\circledast X\in \mathbb{Z}^{b\times (H-R+1)\times (W-R+1)}\\
        &Y[i,j,k]=\sum_{l=0}^{b-1}\sum_{m=0}^{R-1}\sum_{h=0}^{R-1}W[i,l,m,h]X[l,j+m,k+h]
    \end{aligned}
\end{equation}
The polynomials $\hat{x}=\pi'_{\mathrm{X}}({X}), \hat{w}=\pi'_{\mathrm{W}}({W})$ after~\encode~are
\begin{equation}
    \begin{aligned}
        & \hat{x}\in \mathbb{A}_n, \hat{x}[iHW+jW+k]=X[i,j,k]\\
        & \hat{w}\in \mathbb{A}_n, \hat{w}[iHW+(W+1)(R-1)-jW-k]=W[i,0,j,k]\\
    \end{aligned}
\end{equation}
The other slots of $\hat{w}$ are set to 0. The polynomial multiplication result $\hat{y}=\hat{w}\times \hat{x}$ directly gives the convolution result $Y$ as
\begin{equation}
    \begin{aligned}
        \hat{y}&=\hat{w}\times \hat{x}\in \mathbb{A}_n\\
        \hat{y}[iHW+(W+1)(R-1)+jW+k]&=\sum_{l=0}^{b-1}\sum_{m=0}^{R-1}\sum_{h=0}^{R-1}\\
        (\hat{w}[(i-l)HW+(W+1)(R-1)-mW-h]&\hat{x}[lHW+(j+m)W+(k+h)])\\
        &=\sum_{l=0}^{b-1}\sum_{m=0}^{R-1}\sum_{h=0}^{R-1}W[i,l,m,h]X[l,j+m,k+h]\\
        &=Y[i,j,k]\\
    \end{aligned}
\end{equation}
Besides, we extend the definition of $\hat{w}[(i-l)HW+\ldots]=\hat{w}[(b+i-l)HW+\ldots],\forall i<l$.




\newpage
\section*{NeurIPS Paper Checklist}

\begin{enumerate}

\item {\bf Claims}
    \item[] Question: Do the main claims made in the abstract and introduction accurately reflect the paper's contributions and scope?
    \item[] Answer: \answerYes{} 
    \item[] Justification: /

\item {\bf Limitations}
    \item[] Question: Does the paper discuss the limitations of the work performed by the authors?
    \item[] Answer: \answerYes{} 
    \item[] Justification: See Section~\ref{sec:limit}.

\item {\bf Theory Assumptions and Proofs}
    \item[] Question: For each theoretical result, does the paper provide the full set of assumptions and a complete (and correct) proof?
    \item[] Answer: \answerYes{} 
    \item[] Justification: Proofs of all theoretical results are available in the Appendix.

    \item {\bf Experimental Result Reproducibility}
    \item[] Question: Does the paper fully disclose all the information needed to reproduce the main experimental results of the paper to the extent that it affects the main claims and/or conclusions of the paper (regardless of whether the code and data are provided or not)?
    \item[] Answer: \answerYes{} 
    \item[] Justification: Our code and checkpoints are available on \href{https://github.com/Tianshi-Xu/PrivCirNet}{Git Hub}.

\item {\bf Open access to data and code}
    \item[] Question: Does the paper provide open access to the data and code, with sufficient instructions to faithfully reproduce the main experimental results, as described in supplemental material?
    \item[] Answer: \answerYes{} 
    \item[] Justification: Our code and checkpoints are available on \href{https://github.com/Tianshi-Xu/PrivCirNet}{Git Hub}.

\item {\bf Experimental Setting/Details}
    \item[] Question: Does the paper specify all the training and test details (e.g., data splits, hyperparameters, how they were chosen, type of optimizer, etc.) necessary to understand the results?
    \item[] Answer: \answerYes{} 
    \item[] Justification: All training details are available in our code on \href{https://github.com/Tianshi-Xu/PrivCirNet}{Git Hub}.

\item {\bf Experiment Statistical Significance}
    \item[] Question: Does the paper report error bars suitably and correctly defined or other appropriate information about the statistical significance of the experiments?
    \item[] Answer: \answerNo{} 
    \item[] Justification: We conduct experiments on multiple models and datasets, which require significant computational resources. In addition, our code has been released, making it easy to reproduce the results.

\item {\bf Experiments Compute Resources}
    \item[] Question: For each experiment, does the paper provide sufficient information on the computer resources (type of compute workers, memory, time of execution) needed to reproduce the experiments?
    \item[] Answer: \answerYes{} 
    \item[] Justification: Information about our computing resources is available in the Appendix.
    
\item {\bf Code Of Ethics}
    \item[] Question: Does the research conducted in the paper conform, in every respect, with the NeurIPS Code of Ethics \url{https://neurips.cc/public/EthicsGuidelines}?
    \item[] Answer: \answerYes{} 
    \item[] Justification: /

\item {\bf Broader Impacts}
    \item[] Question: Does the paper discuss both potential positive societal impacts and negative societal impacts of the work performed?
    \item[] Answer: \answerNA{} 
    \item[] Justification: /
    
\item {\bf Safeguards}
    \item[] Question: Does the paper describe safeguards that have been put in place for responsible release of data or models that have a high risk for misuse (e.g., pretrained language models, image generators, or scraped datasets)?
    \item[] Answer: \answerNA{} 
    \item[] Justification: /

\item {\bf Licenses for existing assets}
    \item[] Question: Are the creators or original owners of assets (e.g., code, data, models), used in the paper, properly credited and are the license and terms of use explicitly mentioned and properly respected?
    \item[] Answer: \answerYes{} 
    \item[] Justification: /

\item {\bf New Assets}
    \item[] Question: Are new assets introduced in the paper well documented and is the documentation provided alongside the assets?
    \item[] Answer: \answerNA{} 
    \item[] Justification: /

\item {\bf Crowdsourcing and Research with Human Subjects}
    \item[] Question: For crowdsourcing experiments and research with human subjects, does the paper include the full text of instructions given to participants and screenshots, if applicable, as well as details about compensation (if any)? 
    \item[] Answer: \answerNA{} 
    \item[] Justification: /

\item {\bf Institutional Review Board (IRB) Approvals or Equivalent for Research with Human Subjects}
    \item[] Question: Does the paper describe potential risks incurred by study participants, whether such risks were disclosed to the subjects, and whether Institutional Review Board (IRB) approvals (or an equivalent approval/review based on the requirements of your country or institution) were obtained?
    \item[] Answer: \answerNA{} 
    \item[] Justification: /

\end{enumerate}


\end{document}